\documentclass[amsmath,amssymb,aps,pra,twocolumn,showpacs,10pt,floatfix]{revtex4-2}
\usepackage{graphicx}
\usepackage{dcolumn}
\usepackage{bm}
\usepackage{dsfont}
\usepackage[utf8]{inputenc}

\usepackage{braket}
\usepackage{bm}

\usepackage{bbold}

\usepackage{amsmath}
\usepackage{amsthm}
\usepackage{xcolor}
\usepackage[makeroom]{cancel}
\usepackage{enumitem}
\usepackage{hyperref}

\newcommand{\Realpart}[1]{{\rm Re}\left\{ #1\right\}}
\newcommand{\Impart}[1]{{\rm Im}\left\{ #1\right\}}
\newcommand{\tr}[1]{{\rm Tr}\left\{ #1\right\}}

\graphicspath{{}{./}}

\begin{document}

\newcommand{\ave}[1]{\langle #1\rangle}
\newcommand{\wv}[3]{{}_{#1}\ave{#2}_{#3}}
\def\re{\mathrm{Re}}
\def\im{\mathrm{Im}}
\newcommand{\expt}[3]{\left<{#1}\left|{#2}\right|{#3}\right>}
\def\pdag{{\phantom{\dagger}}}
\newcommand{\rs}{\rm \scriptscriptstyle}
\def\alphare{\alpha_{\re}}
\def\alphaim{\alpha_{\im}}
\def\betare{\beta_{\re}}
\def\betaim{\beta_{\im}}
\newcommand\mdoubleplus{\mathbin{+\mkern-10mu+}}

\renewcommand{\CancelColor}{\color{red}}

\def\mathunderline#1#2{\color{#1}\underline{{\color{black}#2}}\color{black}}

\renewcommand{\Re}{\Realpart}

\renewcommand{\Im}{\Impart}

\definecolor{MyGreen}{RGB}{76,203,135}

\title{Distinctive class of dissipation-induced phase transitions and their universal characteristics}

\date{\today}

\author{Matteo Soriente}
\affiliation{Institute for Theoretical Physics, ETH Zurich, 8093 Z{\"u}rich, Switzerland}
\author{Toni L. Heugel}
\affiliation{Institute for Theoretical Physics, ETH Zurich, 8093 Z{\"u}rich, Switzerland}
\author{Keita Arimitsu}
\affiliation{Institute for Theoretical Physics, ETH Zurich, 8093 Z{\"u}rich, Switzerland}
\author{R. Chitra}
\affiliation{Institute for Theoretical Physics, ETH Zurich, 8093 Z{\"u}rich, Switzerland}
\author{Oded Zilberberg}%
\affiliation{Institute for Theoretical Physics, ETH Zurich, 8093 Z{\"u}rich, Switzerland}

\begin{abstract}

Coupling a system to a nonthermal environment can profoundly affect the phase diagram of the closed system, giving rise to a special class of dissipation-induced phase transitions. Such transitions take the system out of its ground state and stabilize a higher-energy stationary state, rendering it the sole attractor of the dissipative dynamics. In this paper, we present a unifying methodology, which we use to characterize this ubiquitous phenomenology and its implications for the open system dynamics. Specifically, we analyze the closed system's phase diagram, including symmetry-broken phases, and explore their corresponding excitations' spectra. Opening the system, the environment can overwhelm the system's symmetry-breaking tendencies, and changes its order parameter. As a result, isolated distinct phases of similar order become connected, and new phase-costability regions appear. Interestingly, the excitations differ in the newly connected regions through a change in their symplectic norm, which is robust to the introduction of dissipation. As a result, by tuning the system from one phase to the other across the dissipation-stabilized region, the open system fluctuations exhibit an exceptional point like scenario, where the fluctuations become overdamped, only to reappear with an opposite sign in the dynamical response function of the system. The overdamped region is also associated with squeezing of the fluctuations.
We demonstrate the pervasive nature of such dissipation-induced phenomena in two prominent examples, namely, in parametric resonators and in light-matter systems. Our work draws a crucial distinction between quantum phase transitions and their zero-temperature open system counterparts.

\end{abstract}

\maketitle

\section{Introduction}

\begin{figure}[t!]
    \includegraphics[width=\columnwidth]{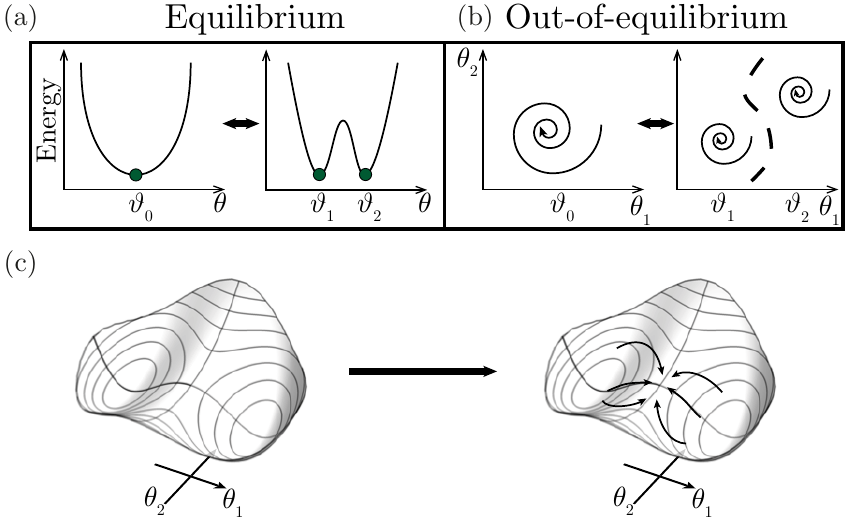}
    \caption{Zero-temperature phase transitions. (a) The Ginzburg-Landau energy functional as a function of a coordinate, $\theta$. A (second-order) $\mathds{Z}_2$ symmetry-breaking phase transitions takes place as model parameters are varied and the ground state of the system changes from having an order parameter $\vartheta_0$ to two degenerate states with order parameters $\vartheta_1$ and $\vartheta_2$. (b) Out-of-equilibrium quantum phase transition in phase space, which is spanned by the coordinates $\theta_{1,2}$. Similar to the equilibrium case, phase transitions (bifurcations) occur when sharp changes occur for the attractors as the model parameters are varied. Crucial to our work, the bifurcations can occur due to a change in the potential energy, but also due to increased dissipative forcing. (c) The system is also characterizable using a (quasi-)energy functional in phase space $(\theta_1,\theta_2)$, where dissipative force terms are introduced via additional arrows, which can stabilize high energy states of the closed system.}
    \label{fig:explain_cartoon}
    \end{figure}

In equilibrium and at zero temperature, we observe so-called quantum phase transitions (PTs), which are generally described through changes in the system's Ginzburg-Landau (GL) energy functional and its symmetries as a control parameter is tuned~\cite{Sachdev_book}, see Fig.~\ref{fig:explain_cartoon}(a). New minima and maxima develop in the GL energy functional when the system's symmetries are broken, and the ground state of the system changes. Correspondingly, the PTs manifest as sharp changes in the system's observables (order parameters), e.g., a second-order PT corresponds to an abrupt change in the derivative of the order parameter (fluctuations). Depending on the specific potential deformation, PTs of different orders occur. 

Phase transitions also transpire in out-of-equilibrium systems, where the interplay between coherent and incoherent dynamics compete in phase space to give rise to new features~\cite{Petruccione_book}. 
Specifically, the system's potential energy is extended to a phase-space energy functional, and environment-induced dissipation channels act as additional forcing terms, see Figs.~\ref{fig:explain_cartoon}(b) and~\ref{fig:explain_cartoon}(c). As a result, open systems are described in terms of their steady-states (stable attractors), where the coherent and incoherent forcing cancel out. Similar to the equilibrium case, dissipative PTs correspond to an abrupt change in order parameters corresponding to the stable attractor bifurcation topology. In this paper, we focus on the situation where the incoherent dissipative channels overwhelm a purely coherent symmetry-breaking PT.

The controllability, offered by numerous contemporary experimental platforms, such as cold atoms~\cite{Ritsch_2013, Li_2019}, trapped ions~\cite{Blatt_2012, Feldker_2020}, superconducting circuits~\cite{Schmidt_2013, Fitzpatrick_2017}, and exciton-polariton cavities~\cite{Carusotto_2013, Fink_2018}, placed out-of-equilibrium PTs at the avantgarde of contemporary research. This has entailed the development of methods and concepts to characterize out-of-equilibrium PTs, ranging from mean-field semiclassical equations of motion (EOM) and their corresponding fluctuations~\cite{Petruccione_book}, to Keldysh action formalism~\cite{Kamenev_book, Sieberer_2016} and third quantization~\cite{Prosen_2008, Prosen_2010}, alongside with the study of exceptional points~\cite{Heiss_2012, ElGanainy_2018} and Liouvillian gaps~\cite{Kessler_2012, Minganti_2018}. Yet, an overarching unifying framework remains missing.

In this paper, we introduce a general framework with which to analyze and understand the ubiquitous characteristics of dissipation-induced PTs. Specifically, we show how dissipation lifts the boundaries of equilibrium (closed) PTs and is able to stabilize states with an energy higher than that of the ground state of the closed system. We identify telltales in the closed system's potential energy and associated fluctuations' spectrum, where dissipation is prone to induce a PT. Specifically, this situation arises at regions where the closed system weakly breaks a symmetry, and the dissipation channel manages to overwhelm the symmetry-breaking PT. The open system then exhibits an exceptional point scenario where the fluctuations become squeezed and overdamped.
Due to the appearance of the dissipation-stablized phase, regions that were distinctly isolated in the closed system phase diagram become connected. These regions have seemingly identical steady states but interestingly differ in their dynamical responses. We highlight the broad applicability of dissipation-induced phenomena by studying how they arise in two prominent out-of-equilibrium systems, namely, the Kerr parametric oscillator (KPO) and the interpolating Dicke-Tavis-Cummings model (IDTC).

Our paper is structured as follows: In Sec.~\ref{sec:closed_open_systems}, we provide a comprehensive overview on the machinery for studying in- and out-of-equilibrium systems via (a) their mean-field solution to obtain the system's fixed points, (ii) analyzing the fixed-points' stability through linearization of the equations of motion near the fixed points, and (iii) Keldysh action formalism for deriving dynamical response functions. In Sec.~\ref{sec:dissiaptive_PT}, we harness the introduced machinery to highlight key signatures of dissipation-induced PTs, and demonstrate its ubiquitous phenomenology in two unrelated paradigmatic examples, namely, the KPO~\cite{Nayfeh_book_Nonlinear_Oscillations,Leghtas_2015, Heugel_2019} and the IDTC model~\cite{Baksic_2014,Soriente_2018}. We conclude and discuss our results in Sec.~\ref{sec:conclusion}.

\section{Characterizing closed and open systems} 
\label{sec:closed_open_systems}
	
Coupling a system to an environment opens it up to dissipation channels and can lead to dissipation-induced PTs, which is the focus of this paper. In such transitions, the incoherent channels can dramatically alter the system's steady state, overruling the closed system's ground-state physics, and forming new regimes defined by stable attractors of the open system. 

In this section, we provide a step-by-step guide of the general framework for analyzing dissipation-induced transitions and their properties; for a more detailed overview and pedagogical approach, we refer the reader to dedicated literature~\cite{Stefanucci_book,Kamenev_book,Weiss_book,Sachdev_book}. At every step, we highlight the relevant assumptions, and in Sec.~\ref{subsec:limitations} discuss the range of applicability of our framework.

In Sec.~\ref{sec:dissiaptive_PT}, we use this formalism to study the impact of dissipation in the presence of a $\mathds{Z}_2$ symmetry breaking. Specifically, we discuss two seemingly unrelated physics problems, a parametrically driven Kerr resonator and a coupled cavity-atom system. We show that in both cases, the dissipation can (i) overwhelm the symmetry breaking, thus (ii) connecting regions in parameter space that are otherwise disconnected, where (iii) an excited state of the system characterized by a negative mass instability is stabilized.

In Sec.~\ref{subsec:closed_sys}, we succinctly review the analysis of ground and excited states of a closed system, based on Landau-Ginzburg energy functionals and fluctuation spectra~\cite{Petruccione_book}. The latter is particularly useful to help identify physical states that can be stabilized by dissipation. In Sec.~\ref{subsec:open_and_steady_states}, we similarly review the analysis of stable and unstable steady states of open systems in appropriate rotating pictures, using mean-field quasi energy functionals (i.e., the energy in a rotating phase-space frame), and dissipative Bogoliubov excitation spectra~\cite{Bogolyubov_1947,Ginzburg_2009}. The latter results from the linear stability analysis of steady states. As we shall see in Secs.~\ref{sec:parametric_kerr_oscillator} and~\ref{sec:IDTC}, in concomitance with a $\mathds{Z}_2$ symmetry-breaking dissipation-induced transition, an exceptional pointlike scenario emerges, i.e., the Bogoliubov spectrum exhibits a pair of complex-conjugated eigenvalues whose imaginary parts go to zero while the real parts split~\cite{Heiss_2012,Ozdemir_2019}. Such behavior similarly marks squeezing in the fluctuations of the system.

Finally, using a Keldysh action approach~\cite{Kamenev_book,Sieberer_2016}, we derive frequency-resolved observables, such as the spectral function [$\mathcal{A}(\omega)$], see Sec.~\ref{subsec:dynamical_observables}. The response functions present peaks corresponding to fluctuation resonances of the system. These resonances correspond to poles that coincide with the dissipative Bogoliubov excitation spectra, described in Sec.~\ref{subsec:open_and_steady_states}.
As we show in Secs.~\ref{sec:parametric_kerr_oscillator} and~\ref{sec:IDTC}, a $\mathds{Z}_2$ symmetry-breaking dissipation-induced transition connects regions in parameter space, with steady states characterized by the same order parameter, but with different dynamical responses. This is traced back to the fact that the states in the disconnected regions of the closed system correspond to ground or excited states, and the dissipation smoothly connects between them.

For pedagogical clarity, we use the example of a simple harmonic oscillator to illustrate our methodology. We refer to the harmonic oscillator Hamiltonian 
\begin{equation}
\label{eq:harmonic_oscillator}
	H = \omega_0\, \hat{a}^\dagger \hat{a} = \frac{1}{2}\begin{pmatrix}
				\hat{a}^\dagger   &   \hat{a}
			\end{pmatrix}\begin{pmatrix}
				\omega_0 & 0 \\ 
				0 & \omega_0		\end{pmatrix}\begin{pmatrix}
				\hat{a}   \\  \hat{a}^\dagger
			\end{pmatrix} - \frac{\omega_0}{2}\,,\, 
\end{equation}
where we set $\hbar=1$ and $\omega_0$ is the characteristic frequency of the oscillator. The operator $\hat{a}$ annihilates an excitation in the oscillator, and we rewrite the Hamiltonian in matrix form for convenience. 

\subsection{Closed system}
	\label{subsec:closed_sys}
	We first consider the equilibrium physics of closed systems, where the Hamiltonian encloses all the information describing the system. It is commonly difficult to solve the Hamiltonian exactly, and we present here a mean-field treatment of the problem and a study of the fluctuations around it.

\begin{figure}[t!]
        \includegraphics[width=\columnwidth]{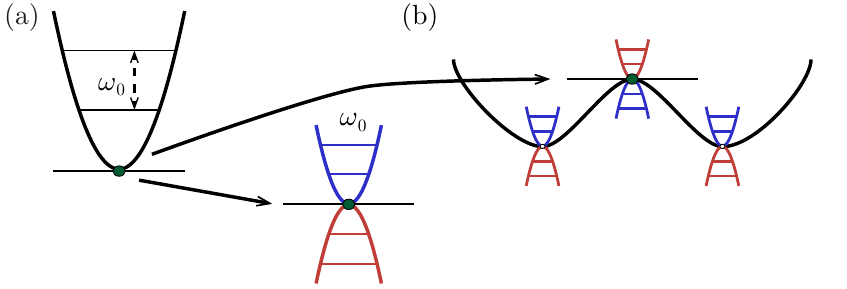}
        \caption{Energy quantization in a parabolic potential. (a) A harmonic potential can also be written as a sum over a particle (blue) and a hole (red) potentials, cf.~Eq.~\eqref{eq:harmonic_oscillator}, corresponding to the two basis solutions of the problem. (b) Changing the potential, a $\mathds{Z}_2$ symmetry-breaking transition can occur, where two minima separated by a maximum appear, each with their own excitation spectrum.}
        \label{fig:ladder}
    \end{figure}

	\textit{Mean-field solutions} --- When studying a system comprised of many degrees of freedom, e.g., an ensemble of harmonic oscillators, we study the mean-field (Landau-Ginzburg) energy functional, defined for $\hat{a}_i\rightarrow  \alpha_i$, where $\alpha_i$ is a complex number corresponding to the semiclassical limit (coherent state) of the operator $\hat{a}_i$, and $i$ iterates over all $N$ degrees of freedom of the system. An inspection of the resulting $N$-dimensional complex functional provides insightful information on the states of the system. Specifically, the extremal points of the functional $\bar{\alpha}_j=\left(\alpha_{1,j},\alpha_{2,j},\ldots\alpha_{N,j}\right)$ correspond to the ground state of the system, as well as to possible excited states. The ground state is defined as the lowest-energy state, see Fig.~\ref{fig:explain_cartoon}(a). For the harmonic oscillator~\eqref{eq:harmonic_oscillator}, the Landau-Ginzburg energy functional is parabolic, $\bar{H} = \omega_0 |\alpha|^2$, with only one global minimum at $\alpha_0=0$, see Fig.~\ref{fig:ladder}(a).

	\textit{Excitation spectrum} --- Having identified the extrema $\hat{\alpha}_j$ of the Landau-Ginzburg energy functional, each of these extrema are characterized by an excitation spectrum defining the fluctuations around the extremal point. This spectrum is obtained via a substitution $\hat{a}_i=\alpha_{i,j}+\delta\hat{a}_{i,j}$ into the original Hamiltonian and by retaining terms up to second order in the fluctuation operators $\delta\hat{a}_{i,j}=\hat{a}_i-\alpha_{i,j}$. Then, using a Bogoliubov transformation, we diagonalize the resulting respective quadratic excitation Hamiltonians, $\hat{H}_{\textrm{exc}}$, see Eq.~\eqref{eq:dynamical_matrix_def} below. The mean-field solutions are physical only if their associated (Bogoliubov) excitation energies are real. 
	Diagonalizing the excitation Hamiltonian, we ensure that the diagonalization procedure preserves the statistics of the excitations. To this end, the proper Bogoliubov transformation~\cite{Stefanucci_book,Xiao_2009,Soriente_2020} for bosonic systems is found by diagonalizing the dynamical matrix,
	\begin{equation}
	\label{eq:dynamical_matrix_def}
		\hat{D}_{\textrm{exc}}\equiv \hat{I}_- \cdot \hat{H}_{\textrm{exc}}\,,
	\end{equation} 
	where $\hat{I}_-=\mathds{1}_{N}\otimes (-\mathds{1}_{N})$ is a diagonal matrix with $+ 1$ ($-1$) entries on the first (second) $N$ elements with $2N\times 2N$ the size of $\hat{H}_{\textrm{exc}}$. Note that the definition of $\hat{I}_-$ relies on the operator ordering in the matrix definition of the Hamiltonian $\hat{H}_{\textrm{exc}}$, cf.~Eq.~\eqref{eq:harmonic_oscillator}. Through the Bogoliubov transformation, we obtain the excitation eigenmodes of the system, characterized by the eigenvalues and eigenvectors of $\hat{D}_{\textrm{exc}}$. The $2N$ excitation frequencies are paired with opposite signs $\pm\omega_i\in\mathds{R}$ ($i=1,\ldots,N$). In the case of the harmonic oscillator Eq.~\eqref{eq:harmonic_oscillator}, the excitation Hamiltonian coincides with the bare Hamiltonian. We diagonalize Eq.~\eqref{eq:harmonic_oscillator}, and find the familiar ladder of equispaced excited states, where the ground state coincides with the lowest excitation eigenstate, see Fig.~\ref{fig:ladder}(a).

	\textit{Norm inversion in the excitations} --- The excitations' eigenvectors further provide us with the symplectic norm~\cite{Soriente_2020}
	\begin{equation}
	\label{eq:symplectic_norm}
		ds_{\mathbf{v}_j}^2\equiv \mathbf{v}_j^\dag \hat{I}_- \mathbf{v}_j\,,
	\end{equation}
	associated to each excitation eigenfrequency, where $\mathbf{v}_j$ with $j=1,\ldots,2 N$ are the eigenvectors of $D_{\textrm{exc}}$. The symplectic norm Eq.~\eqref{eq:symplectic_norm} determines the nature of the excitations: It can be either positive ($ds^2>0$) or negative ($ds^2<0$) and is a measure of the particle- or hole-like nature of the excitation, respectively~\cite{Bogolyubov_1947,Stefanucci_book,Soriente_2020}. 	
	For instance, the ground state of the system as well as other local minima of the energy functional are accompanied by particlelike (holelike) eigenmodes at positive (negative) frequencies, whereas local maxima are associated with a holelike (particlelike) eigenmode at positive (negative) frequencies. In Fig.~\ref{fig:ladder}(a), we sketch the particle- (blue) versus holelike (red) ladders for the harmonic oscillator excitations around the ground state.
	The distinction between particle- and holelike modes is crucial and it assists us in identifying excited states that correspond to maxima in the mean-field energy functional, which can become a stable steady state due to a dissipation-induced PT.
	
	As we tune an external parameter, the shape of the mean-field energy functional can change and develop new features. For example, a (second-order) spontaneous $\mathds{Z}_2$ symmetry-breaking PT can occur when the energy landscape moves from exhibiting a single ground state to two degenerate ground states separated by a maxima, see Fig.~\ref{fig:ladder}(b). In such a scenario, the symmetry-conserving mean-field solution now becomes an excited state. As the system traverses the critical point, the symplectic norm of the excited state changes signs. In the closed system setting, this state cannot be reached by adiabatic ground state evolution; we will show later that dissipation can stabilize such a steady state.

	 \begin{figure}[t!]
        \includegraphics[width=\columnwidth]{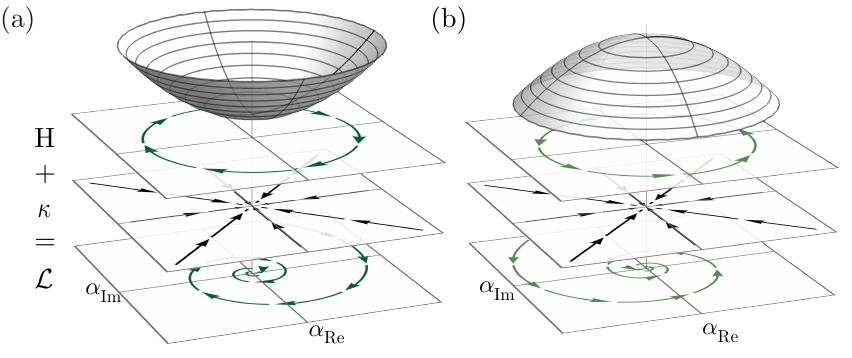}
        \caption{Dissipative harmonic oscillator, cf.~Eqs.~\eqref{eq:harmonic_oscillator} and \eqref{eq:Lindblad_kappa}. (a), (b) The rotating mean-field energy potential landscape $\bar{H}_{\textrm{rot}}$ (cf.~Sec.~\ref{subsec:static_HO}) for negative and positive detuning $\Delta$. The arrows indicate the resulting Hamiltonian motion. When $\Delta$ changes its sign, the peak in $\mathcal{A}_{\textrm{rot}}$ passes $0$, $\bar{H}_{\textrm{rot}}$ changes from a valley to a hill, and the rotation around the origin (center) changes its direction. Adding dissipation (arrows) leads to a flow toward the empty state at the center.}
        \label{fig:HO_H+L}
    \end{figure}

\subsection{Open system}
\label{subsec:open_and_steady_states}

To characterize the physics of open systems, we first find the so-called steady (or stationary) states, which correspond to the long-time behavior of the open system. To do so, we solve Liouville's master equation~\cite{Petruccione_book},
	\begin{equation}
	    \label{eq:master_eq}
		\frac{d \rho}{dt} = L [\rho]\,,
	\end{equation}
	in the long-time limit. Here $\rho$ is the density matrix of the system and $L$ is the Liouvillian superoperator. We obtain the steady states by requiring that the system's density matrix does not change in time, i.e., $d\rho/dt \equiv 0$. Note that rotating steady states, e.g., limit cycles, fulfill a similar condition in the correct rotating frame. 
    
    In this paper, we consider cases with a strong separation of time scales between system and environment, and assume a memoryless environment with weak system-environment coupling, such that we can perform the Born-Markov and secular approximations, thus simplifying Eq.~\eqref{eq:master_eq} to obtain Lindblad's master equation~\cite{Gardiner_book_Quantum_Noise,Weiss_book,Ferguson_2020_open}:
    \begin{equation}
	\label{eq:Lindblad_master_eq}
		\frac{d \rho}{dt} = -\frac{i}{\hbar}\left[\hat{H}_{\textrm{sys}},\rho\right] + \sum_\alpha \gamma_\alpha\mathcal{L}[\hat{L}_\alpha]\rho\,.
	\end{equation}
    The unitary evolution generated by the system's Hamiltonian is described through the commutator, while the dissipative dynamics with rates $\gamma_\alpha$ appear in Lindblad form,
    \begin{equation}
		\mathcal{L}[\hat{L}_\alpha]\rho = 2\hat{L}_\alpha \rho \hat{L}_\alpha^\dagger - \left\{\hat{L}_\alpha^\dagger \hat{L}_\alpha,\rho\right\}\,,
		\label{eq:Lindblad_dissipator}
	\end{equation}
	where the Lindblad dissipator is defined in terms of so-called Lindblad operators (or quantum jump operators) $\hat{L}_\alpha$ that describe the coupling to the environment. The anticommutator term corresponds to dissipation, while the so-called \emph{recycling} or \emph{quantum jump} term, $2\hat{L}_\alpha \rho \hat{L}_\alpha^\dagger$, encodes fluctuations and ensures the normalization $\tr{\rho} \equiv 1$ of the system's density matrix.

	In the context of our pedagogical example Eq.~\eqref{eq:harmonic_oscillator} and throughout this paper, dissipation due to coupling to a zero temperature environment takes the Lindblad form
	\begin{equation}
	    \label{eq:Lindblad_kappa}
         \mathcal{L}[\hat{a}]\rho = \kappa [2\hat{a} \rho \hat{a}^\dagger - \{\hat{a}^\dagger \hat{a},\rho\}]\,,
    \end{equation}
    where $\kappa\ll \omega_0$ is the positive dissipation rate describing loss of excitations. We can pictorially describe the harmonic oscillator evolution under Eq.~\eqref{eq:Lindblad_master_eq}. The eigenstates of the closed system describe closed circular orbits along equipotential lines of the Landau-Ginzburg energy functional. We dress the Landau-Ginzburg energy functional with arrows whose directions depend on the particular dissipation channel. As the dissipation Eq.~\eqref{eq:Lindblad_kappa} destroys excitations in the system, the arrows point toward the empty resonator state. In other words, the dissipation pushes the closed system orbits inwards. The net result is a spiraling evolution toward the center, which is the sole steady state for the dissipative harmonic oscillator, see Fig.~\ref{fig:HO_H+L}(a).

    \textit{Mean-field steady states} --- The master equations, Eqs.~\eqref{eq:master_eq} and~\eqref{eq:Lindblad_master_eq}, evolve the state of the system. To better characterize its steady state, we use instead the time evolution of mean-field expectation values in Schr{\"o}dinger's picture,
	\begin{equation}
	\label{eq:time_evo_operator}
	    \frac{d\ave{\hat{o}}}{dt} = \tr{\hat{o}\frac{d\rho}{dt}}\,,
	\end{equation}
where $\ave{\hat{o}}$ is the expectation value of one of the observable operators describing the system. Considering different observables, we obtain a set of coupled first-order differential equations,
	\begin{equation}
	\label{eq:EOM_matrix}
		\frac{d\mathbf{O}}{dt} = M \mathbf{O}\,,
	\end{equation}
	where $\mathbf{O} = (\ave{\hat{o}_1},\ave{\hat{o}_2},\ave{\hat{o}_3},...)$ is the basic set of real operator expectation values describing our system. Note that for bosonic coherent states $\alpha=\ave{\hat{a}}$ is a complex number and we split its evolution into its real and imaginary parts in Eq.~\eqref{eq:EOM_matrix}. Taking the steady state, $d O/dt \equiv 0$, we solve the resulting set of algebraic equations with multiple solutions that satisfy $M O = 0$. We find the set of steady-state solutions and henceforth denote by $O_{i}^j$ the expectation values corresponding to the $j$th physical solution of the $i$th operator. Note that physical solutions have real observables, i.e., $\Impart{O_{i}^j}=0$, whereas Eqs. set~\eqref{eq:EOM_matrix} can yield unphysical (complex) solutions.

	\textit{Stability analysis} --- Among the physical steady states, some are stable and some are unstable against linear fluctuations, cf.~the above discussion on excitations in the closed system. Depending on the initial boundary conditions, the system will eventually evolve towards one of the former. To find which steady states are stable or not, we perform a linear expansion of Eq.~\eqref{eq:EOM_matrix} around each solution according to $\hat{o}_{i} = O_{i}^j + \delta\hat{o}_{i}^j$, where $\delta\hat{o}_{i}^j\equiv \hat{o}_{i} - O_{i}^j$ is a fluctuation operator around $O_{i}^j$. We then define the stability matrix $M^j$, whose entries depend on the $j$th solution, and which defines the fluctuation dynamics around this solution through
	\begin{equation}
	\label{eq:fluc_stability}
		\frac{d{\delta \hat{\mathbf{O}}}^j}{dt} = M^j {\delta \hat{\mathbf{O}}}^j\,,
	\end{equation}
	where ${\delta \hat{\mathbf{O}}}^j = (\ave{\delta\hat{o}_{1}^j},\ave{\delta\hat{o}_{2}^j},\ave{\delta\hat{o}_{3}^j},\ldots)$ is the set of the fluctuation operators' expectation values describing our system.
	Equation~\eqref{eq:fluc_stability} defines a set of coupled linear differential equations, where the eigenvalues $\epsilon_i$ of the stability matrix $M^j$ determine the stability of the $j$th solution. This results from the fact that the eigenmodes of $M^j$ evolve in time according to $e^{\epsilon t}$. Then, the real part of $\epsilon$ plays the role of a lifetime of the excitation mode, whereas the imaginary part is its frequency. In other words, if at least one eigenvalue $\epsilon$ has a positive real part, the fluctuations diverge and the solution is unstable.
	
	\textit{Variance} --- In similitude to equilibrium statistical physics PTs in open systems can be classified through discontinuities in the properties of the order parameters and their fluctuations~\cite{Sachdev_book}. Specifically, first-order PTs manifest in a discontinuous change in the mean-field order parameters, second-order PTs show a discontinuity in the fluctuations of the system, whereas third-order PTs exhibit a discontinuity in the variance of the fluctuations. In the following sections, we show that the dissipation-induced PT connects between two steady states, which were disconnected in the closed system limit, but are characterized by the same mean-field order parameter. Moving between the two newly connected states, a discontinuity in the system's dynamical behavior appears but no discontinuity manifests--neither at the level of fluctuations nor for the variance of the fluctuations.

To show this, in the following, we calculate the covariance of the fluctuations using Eq.~\eqref{eq:time_evo_operator} for the time evolution of the expectation of correlation operators:
\begin{align}
\hat{\mathbf{K}}_{ij} = \delta \hat{o}_i\delta \hat{o}_j\,.
\label{eq:variance_gen}
\end{align}
Taking the lowest-order (mean-field) results for the covariance operators yields systems of coupled equations analogous to Eq.~\eqref{eq:EOM_matrix}, which we solve for the steady-state.

\subsection{Static observables of the harmonic oscillator}
\label{subsec:static_HO}

	 \begin{figure*}[t!]
        \includegraphics[width=\textwidth]{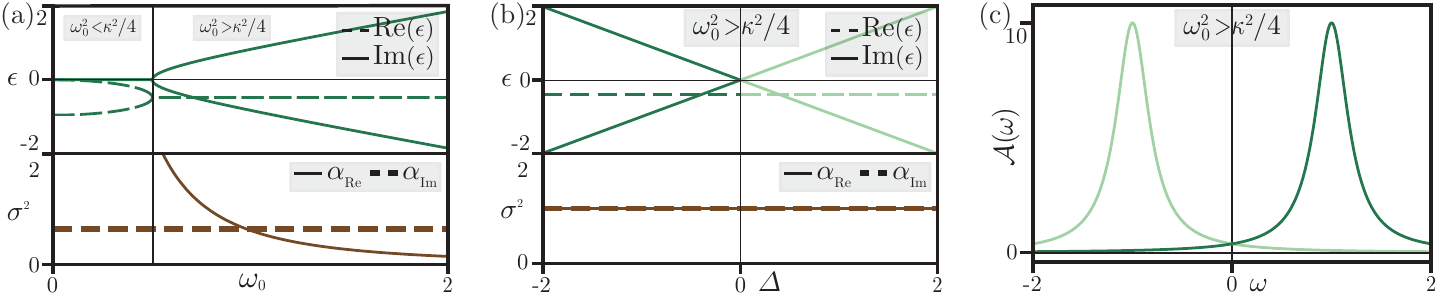}
        \caption{Analysis of the damped harmonic oscillator, cf.~Eqs.~\eqref{eq:harmonic_oscillator} and \eqref{eq:Lindblad_kappa}. (a), (b) Real (dashed) and imaginary (solid) parts of the fluctuation eigenvalues Eq.~\eqref{eq:fluc_stability} (green). The variance of fluctuations Eq.~\eqref{eq:variance_gen} (brown) along the real (dashed) and imaginary (solid) quadratures of the harmonic oscillator. (a) Keeping the dissipation constant and tuning the oscillator's eigenfrequency leads to an overdamped regime, when $\omega_0^2 < \kappa^2/4$. Correspondingly, the eigenvalues show a mode softening but no unstable behavior, i.e., the imaginary parts go to zero for $\omega_0=\kappa/2$, while the real parts always remain negative but split from the value $-\kappa$. The variance associated with the real part is flat along the entire region, while the imaginary part variance increases but remains finite. (b) Same as (a) for the harmonic underdamped oscillator in the rotating frame. In the weak dissipation limit, as a function of detuning $\Delta$, the imaginary part of the fluctuation eigenvalues inverts at $\Delta=0$, while the real parts remain negative and unchanged. The variance remains unchanged, marking the fact that no phase transition is occurring. (c) Underdamped harmonic oscillator response (spectral function) [cf.~Eq.~\eqref{eq:spectral_function}] for positive (light green) and negative (dark green) detunings.}
    \label{fig:HO_var_eigen_resp}
    \end{figure*}

We are now ready to apply our aforementioned machinery onto the dissipative harmonic oscillator introduced in Eqs.~\eqref{eq:harmonic_oscillator} and~\eqref{eq:Lindblad_kappa}. We will explore both the overdamped ($\omega_0^2<\frac{\kappa^2}{4}$) and underdamped ($\omega_0^2> \frac{\kappa^2}{4}$) regimes, as they allow us to introduce key signatures that will be useful for our discussion in Secs.~\ref{sec:parametric_kerr_oscillator} and \ref{sec:IDTC}.
		
\textit{Overdamped oscillator} ---
In the overdamped limit, dissipation can overwhelm the Hamiltonian potential of the system. Note that this limit involves a system-environment coupling that goes beyond the Lindblad limit discussed above~\cite{Pop_1988,Yeon_2001}. Nevertheless, a similar mean-field, stability, and variance analysis as in Eqs.~\eqref{eq:EOM_matrix} and~\eqref{eq:fluc_stability} holds here, see Appendix~\ref{sec:HO_overdamped_appendix}. There is a single steady-state to the system, namely, an inert oscillator. In Fig.~\ref{fig:HO_var_eigen_resp}(a), we show the fluctuation eigenvalues and variance obtained on top of the empty state. We observe that at a certain point, the eigenvalues show an abrupt change (akin to an exceptional point~\cite{Ozdemir_2019}) in their derivative with respect to the control parameter (here the mode's eigenfrequency). Specifically, the imaginary parts become degenerate, while the real parts split in a region in parameter space. This marks the fact that the fluctuations become overdamped and do not exhibit an oscillating behavior. Instead, two characteristic decay times appear [dashed lines in Fig.~\ref{fig:HO_var_eigen_resp}(a)], and lead to squeezing of the fluctuations. Note that the sudden change in the fluctuations eigenvalues does not result in a discontinuous variance, ruling out the presence of a third-order PT. 

\textit{Underdamped resonator} --- In open systems, we often deal with the presence of an external drive. As the system is expected to lock to the drive frequency in the long-time limit, a useful method to treat external drives is to move to a rotating frame with respect to the drive. If the oscillator is driven at a frequency $\omega_{\textrm{drive}}$, with the unitary transformation $T = e^{-i\omega_{\textrm{drive}}t a^\dagger a}$ we move to a rotating frame at frequency $\omega_{\textrm{drive}}$, yielding the rotating frame Hamiltonian $H_{\textrm{rot}} = - \Delta a^\dagger a$ with detuning $\Delta = \omega_{\textrm{drive}} - \omega_{0}$. The sign of the detuning determines whether the drive stiffens or softens the harmonic potential, and additionally, whether the mean-field energy potential landscape, $\bar{H}_{\textrm{rot}} = -\Delta |\alpha|^2$, is a paraboliod with a minimum or a maximum around the phase-space origin, for negative or positive $\Delta$, respectively [see Figs.~\ref{fig:HO_H+L}(a) and~\ref{fig:HO_H+L}(b)]. 
Depending on the detuning, the empty cavity state appears in the rotating frame, either as the ground state (negative detuning) or as an effective excited state (positive detuning). Additionally, the change in curvature of $\bar{H}_{\textrm{rot}}$ as a function of the detuning manifests as the transformation of particle- to hole-like excitations and vice versa through a change in the sign of the symplectic norm Eq.~\eqref{eq:symplectic_norm}. The switch in the norm corresponds to clockwise (counterclockwise) rotations of the state of the system relative to the rotating reference frame (clock), see Fig.~\ref{fig:HO_H+L}.

The effect of dissipation remains unchanged in the rotating frame picture and decreases the amplitude of oscillations, namely, it always creates a dragging force toward the center. The joint effect of coherent and incoherent forces leads to a spiraling evolution of the state of the system toward the minimum (maximum) of the potential in the case of negative (positive) detuning, see Fig.~\ref{fig:HO_H+L}. 
The quasi-energy potential inversion also manifests in the open system fluctuation spectrum in the form of a degeneracy point, where a pair of complex conjugated fluctuation eigenvalues, $\epsilon_i$, acquires a zero imaginary part while the real parts remain negative. The latter condition implies that the system's steady state remains stable at $\Delta = 0$ despite the fact that the effective confining Hamiltonian potential vanishes. At the same time, the variance as a function of detuning $\Delta$ remains unchanged, marking the fact that the degeneracy point at $\Delta = 0$ does not imply a PT, see Fig.~\ref{fig:HO_var_eigen_resp}(b) and Eqs.~\eqref{eq:EVKPO}-\eqref{eq:varKPO2} in Appendix~\ref{sec:KPO_open_appendix} with $G=U=0$.

\subsection{Dynamical observables}
\label{subsec:dynamical_observables}

In the closed system, the sign of the fluctuations' symplectic norm Eq.~\eqref{eq:symplectic_norm} helps us to distinguish between the ground state (positive norm) and an unstable excited state (negative norm) even if the two states are characterized by the same mean-field observable expectation values. In the open system scenario, we do not have the notion of a symplectic norm. We resort instead to study dynamical observables of the system to fully understand the nature of a steady state.

To study dynamical observables, we use a Keldysh action formulation, which for a bosonic theory takes the form~\cite{Kamenev_book,Sieberer_2016}
\begin{equation}
\label{eq:rotated_Keldysh_action}
		S = \int_{-\infty}^{\infty}dt \begin{pmatrix} a_c^* & a_q^* \end{pmatrix}\begin{pmatrix} 0 & {\big[G^A\big]}^{-1} \\ {\big[G^R\big]}^{-1} & D^K \end{pmatrix}\begin{pmatrix} a_c \\ a_q \end{pmatrix}\,,
\end{equation}
where the fields $a_c, a_q$ are the standard classical and quantum bosonic fields acting on the Keldysh contour. This terminology signals that the former combination of fields can acquire a (classical) field expectation value, while the latter one cannot. The details of the derivation can be found in Ref.~\cite{Sieberer_2016}. The strength of this formalism lies in its matrix structure, where the action Eq.~\eqref{eq:rotated_Keldysh_action} is directly connected to the Green's functions of the system, $G^R(t,t')=-i\theta(t-t')\langle[a(t),a^\dagger(t')]\rangle$, $G^A = [G^R]^\dagger$, $D^K$, which are the retarded and advanced Green's functions, as well as the Keldysh component, respectively.
It is useful to move to Fourier space and obtain a frequency-dependent action,
\begin{equation}
\label{eq:rotated_frequency_Keldysh_action}
		S = \int_\omega \begin{pmatrix} a_c^* & a_q^* \end{pmatrix}\begin{pmatrix} 0 & {[G^A]}^{-1}(\omega) \\ {[G^R]}^{-1}(\omega) & D^K(\omega) \end{pmatrix}\begin{pmatrix} a_c \\ a_q \end{pmatrix}\,,
\end{equation}
where $\int_\omega = \int\frac{d\omega}{2\pi}$.

Taking, for example, the harmonic oscillator Eq.~\eqref{eq:harmonic_oscillator}, we can write the action Eq.~\eqref{eq:rotated_frequency_Keldysh_action} as
\begin{equation}
    \label{eq:HO_action}
    \resizebox{.88\hsize}{!}{$\displaystyle{ S = \int_t \begin{pmatrix} a_c^* & a_q^* \end{pmatrix}\begin{pmatrix} 0 & \omega - \omega_0 - i\kappa\ \\ \omega - \omega_0 + i\kappa\ & 2i\kappa \end{pmatrix}\begin{pmatrix} a_c \\ a_q \end{pmatrix}\,,}$}
\end{equation}
where we assumed a Lindlad dissipator in the form of Eq.~\eqref{eq:Lindblad_kappa}. We identify the inverse Green's functions and the Keldysh component with
\begin{align}
	[G^R]^{-1}(\omega) & = {[G^A]^{-1}}^* (\omega) = \omega - \omega_0 + i\kappa\,, \notag \\
	D^K(\omega) & = 2i\kappa\,. \notag
\end{align}
Inverting the matrix action Eq.~\eqref{eq:rotated_Keldysh_action}, we obtain direct access to the frequency-resolved Green’s functions of the system; for the harmonic oscillator Eq.~\eqref{eq:harmonic_oscillator} 
\begin{align}
	G^R(\omega) & = G^A(\omega)^* = \frac{1}{\omega - \omega_0 + i\kappa }\,,\notag \\
	G^K(\omega) & = - G^R(\omega) D^K(\omega) G^A(\omega) = -\frac{2i\kappa}{(\omega - \omega_0)^2 + \kappa^2}\,. \notag
\end{align}

Using these Green’s functions, we can calculate frequency-dependent observable quantities (\textit{dynamical order parameters}), such as the response and correlations of our system. Here, we provide a short list of those correlations that are most-commonly probed:
\begin{enumerate}[leftmargin=*]
	\item The \emph{spectral function} (density) ($\mathcal{A}$).

	In linear response theory, the spectral function describes the excitation that the system undergoes when adding a single particle with frequency $\omega$ to it. It is defined as 
	\begin{equation}
	\label{eq:spectral_function}
		\mathcal{A}(\omega) = i[G^R(\omega) - G^A(\omega)] = -2\Im{G^R(\omega)}\,,
	\end{equation}
	where the retarded Green's function, $G^R(\omega)$, describes the linear response of a system to a weak external perturbation. For bosonic systems, $\mathcal{A}(\omega)$ fulfills the sum rule:
	\begin{equation}
		\int_\omega A(\omega) = \langle [a,a^\dagger]\rangle = 1\,. \notag
	\end{equation}

	\item The \emph{power spectrum} ($\mathcal{C}$).
	
	The power spectrum describes the occupation of the individual quantum excitation modes of the system. It is defined as the Fourier transform of the correlation function $\langle\{a(t),a^\dagger(0)\}\rangle$, yielding
	\begin{equation}
	\label{eq:power_spectrum}
		\mathcal{C}(\omega) = iG^K(\omega)\,,
	\end{equation}
	where $G^K(t,t')=-i\langle\{a(t),a^\dagger(t')\}\rangle$. Considering the steady-state solution, and taking the equal-time limit, we can find the \emph{mode occupation} via
	\begin{equation}
		\label{eq:mode_occupation}
		\langle a^\dagger a\rangle = \frac{1}{2}(i\int_\omega G^K(\omega) - 1)\,.
	\end{equation}
	
	\item The \emph{fluorescence spectrum} ($\mathcal{S}$).
	
	The autocorrelation of the systems' excitations describes the probability of measuring an excitation of frequency $\omega$ leaving the system, dubbed fluorescence spectrum. It is defined as
	\begin{equation}
	\label{eq:fluorescence_spectrum}
		\resizebox{.88\hsize}{!}{$\displaystyle{\mathcal{S}(\omega) = \langle a^\dagger(\omega) a(\omega)\rangle = \frac{i}{2}[G^K(\omega) - G^R(\omega) + G^A(\omega)]\,.}$}
	\end{equation}
	
\end{enumerate}
Each one of these functions presents peaks of different heights at the resonance frequencies of the system under study. In the simple case of the dissipative harmonic oscillator, we find that the spectral function exhibits the familiar line shape peaked at the resonance frequency of the oscillator:
\begin{equation}
	\mathcal{A}(\omega) = \frac{2\kappa}{(\omega - \omega_0)^2 + \kappa^2}\,. \notag
\end{equation}
Interestingly, the mode occupation $\mathcal{C}(\omega)$ in this system coincides with the spectral function as can be seen by comparing $\mathcal{A}(\omega)$ and $G^K(\omega)$. This will not be the case in more complicated systems.

To conclude this section, we consider again the harmonic oscillator in a rotating frame due to the presence of an external drive. The response of the oscillator can be found in the same fashion as before with the only difference that now the resonance frequency is the detuning from the drive
\begin{equation}
    	\mathcal{A}_{\textrm{rot}}(\omega) = \frac{2\kappa}{(\omega + \Delta)^2 + \kappa^2}\,. \notag
    \end{equation}
This implies that the oscillator response is peaked at positive frequencies for negative $\Delta$ and at negative frequencies for positive $\Delta$, as shown in Fig.~\ref{fig:HO_var_eigen_resp}(c). In this example, the empty ($\alpha = 0$) steady state has a different dynamical behavior depending on the sign of $\Delta$, where the two different regions are separated at $\Delta=0$. Two key elements emerge from this simple discussion. First, the steady-state of the system does not change, i.e., it is always the empty state with no excitations in the system. Second, as we tune the external parameter $\Delta$, the dynamical response response shifts from positive to negative frequencies.

\subsection{Limitations and applicability}
\label{subsec:limitations}

The framework we outlined in this section involved certain assumptions, namely (i) we treat bosonic systems or systems that are mappable to bosonic ones, e.g., via the Holstein-Primakoff transformation~\cite{Holstein_1940}, (ii) we employ a mean-field (saddle-node) treatment of the order the parameter of the system on top of which quadratic fluctuations appear, and (iii) we truncate strong correlation between the system and environment (Lindblad approximation), where the environment is taken to be at zero temperature. Concerning (i), our analysis can be readily extended to fermionic or mixed systems, where the particles' commutation relations will impact the specific form of the derived expressions. The assumption (ii) is well justified whenever the system is well within a phase. Expanding beyond mean-field such that higher-order correlations are taken into account can play a crucial role whenever the system is close to criticality. Recently, several approaches have been developed to study higher-order correlations, including dynamical mean-field theory, diagrammatic expansions, functional renormalization group, exact diagonalization, and numerical or density matrix renormalization group analysis~\cite{Bulla_2003,Schollwock_2005,Sieberer_2013,Aoki_2014,Tauber_2014,Mascarenhas_2015,Jin_2016,Mathey_2020,Arndt_2021}. The approach presented in this work can be in principle extended to include higher order fluctuations. Going beyond (iii) can lead to interesting strongly-correlated states between system and environment, e.g., the Kondo effect~\cite{hewson1997kondo}. Systematic diagrammatic expansion of the system environment coupling is challenging and depends crucially on the environment memory, size, and coherence, see e.g., Ref.~\cite{Ferguson_2020_open} and discussion therein. We limit ourselves here to the simplest (and ubiquitous to light-matter systems) case to highlight the profound impact that even such a simple environment has on out-of-equilibrium PTs.

\section{Dissipation-induced phase transitions}
\label{sec:dissiaptive_PT}

Following the pedagogical introduction, we are at a vantage point to better define dissipation-induced PTs. Generally, this includes any scenario where incoherent terms overwhelm coherent Hamiltonian terms and lead to a change in the steady-state topology in phase space, i.e., to bifurcations. Specifically, we now focus on the case where the Hamiltonian prompts a spontaneous $\mathds{Z}_2$ or $U(1)$ symmetry-breaking [cf.~Figs.~\ref{fig:explain_cartoon} and~\ref{fig:ladder}]. Dissipation can overwhelm the expected symmetry breaking and stabilize non-symmetry-broken states. A transition into this region manifests in the form of an exceptional pointlike scenario for the excitation spectrum and squeezing of the variance, as discussed in Fig.~\ref{fig:HO_var_eigen_resp}(a). More importantly, the dissipation-stabilized region bridges between two disconnected parts of the closed system phase diagram. As one traverses these regions, imaginary parts of some modes in the fluctuation spectrum invert, as discussed in Figs.~\ref{fig:HO_var_eigen_resp}(a) and~\ref{fig:HO_var_eigen_resp}(b). Correspondingly, the dynamical response functions display inversions in some of their peaks. In the following, we discuss two seemingly unrelated paradigmatic driven-dissipative systems that exhibit such phenomenology. Thus, we highlight the ubiquitous nature and characteristics of such transitions.


\subsection{Parametric Kerr oscillator} 
\label{sec:parametric_kerr_oscillator}

The first exemplary system we consider is the Kerr oscillator driven by a parametric (two-photon) pump, while subject to single-photon dissipation~\cite{Gibbs_1976, Drummond_1980, Rempe_1991, Casteels_2016, Casteels_2017, Heugel_2019,Sahoo_2020}. 
As a function of the two-photon drive, this KPO exhibits a continuous, $\mathds{Z}_2$ time-translation symmetry-breaking transition from an empty cavity state, dubbed normal phase (NP), to a state with finite photon number, dubbed parametric phase state (PPS)~\cite{Bartolo_2016}.
The symmetry-broken PPSs appear in pairs of coherent states with equal amplitude but $\pi$-shifted phase~\cite{Puri_2017}.
At low photon numbers, their superpositions form Schr\"odinger cat states of opposite parities~\cite{Leghtas_2015, Minganti_2016, Elliot_2016}. 
Possible applications of such phase states include annealing-based optimization algorithms in classical and quantum KPO networks~\cite{Inagaki_2016, Nigg_2017, frimmer2019rapid}, as well as universal quantum computation~\cite{Goto_2016, Puri_2017}.
Furthermore, the competition between single- and two-photon drives leads to PTs that can be used for sensing~\cite{Papariello_2016, Leuch_2016, eichler2018parametric, Heugel_2019}. 
At weak two-photon drives, the dissipation stabilizes the KPO, while squeezing its fluctuations, with various applications in sensing~\cite{Yurke_1988, Castellanos-Beltran_2008,Mohammadi_2020}.
We, now, turn to review how the single-photon dissipation changes the phase diagram of the KPO according to the dissipation-induced phenomenology described above, cf.~Ref.~\cite{Zerbe_1994}.

\subsubsection{Closed system}
\label{subsec:KPO_closed}

We first consider the (closed system, non-dissipative) KPO, with Hamiltonian
\begin{equation}
	\label{eq:KO_Hamiltonian}
	H_{\mathcal{K}} = \omega_c a^\dagger a + \frac{U}{2}a^\dagger a^\dagger a\, a +H_{G}\,,
\end{equation}
where $a$ annihilates a bosonic particle on the oscillator, and $U$ is the Kerr nonlinearity amplitude. A parametric two-photon coherent pump of strength $G$ and frequency $\omega_{G}^{\phantom{\dagger}}$ is described by the Hamiltonian
\begin{equation}
	H_{G} = \frac{G}{2}e^{-i\omega_{G}^{\phantom{\dagger}} t}a^\dagger a^\dagger + \frac{G^*}{2}e^{i\omega_{G}^{\phantom{\dagger}} t}a\,a\,.
\end{equation}
We use the unitary transformation $T = e^{-i a^\dagger a \omega_{G}^{\phantom{\dagger}} t/2}$ to move to a rotating frame with respect to half the parametric drive frequency. Neglecting fast oscillating terms, we obtain the time-independent KPO Hamiltonian in the rotating wave approximation
\begin{equation}
	\label{eq:KPO_Hamiltonian}
	\tilde{H}_{\mathcal{K}} = -\Delta a^\dagger a + \frac{U}{2}a^\dagger a^\dagger a \,a + \frac{G}{2} a^\dagger a^\dagger + \frac{G^*}{2} a \,a\,,
\end{equation}
where $\Delta = \omega_{G}^{\phantom{\dagger}}/2 - \omega_c$ is the half-pump cavity detuning. In the following, we will focus on positive $U$, but similar physics is obtained for negative $U$, using the transformation $\Delta \rightarrow -\Delta$ and $ G \rightarrow -G$.

\begin{figure}[t!]
    \includegraphics[width=\columnwidth]{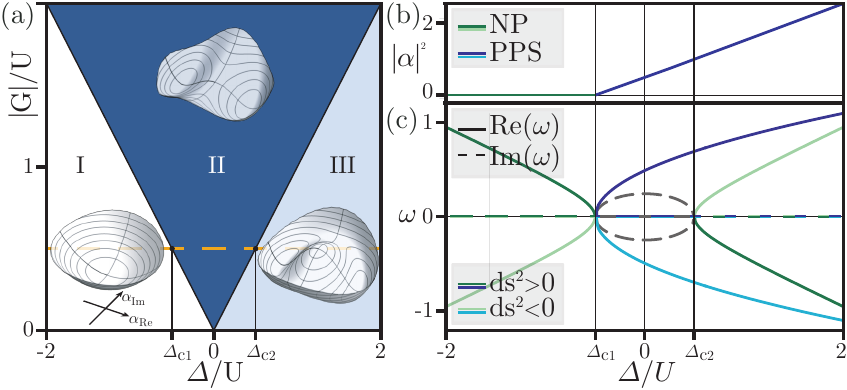}
    \caption{Closed Kerr parametric oscillator [cf.~Eq.~\eqref{eq:KPO_Hamiltonian}]. (a) Phase diagram: The system displays a $\mathds{Z}_2$ symmetry; three distinct regions are indicated by their respective mean-field energy potential landscape [cf.~Eq.~\eqref{eq:KPO_MF_landscape}], as a function of the real and imaginary parts of the cavity field, $\alphare$ and $\alphaim$, respectively. Phase transitions occur between a normal phase (white) and a symmetry broken phase (blue). The light-blue region indicates the parameter regime, where the normal phase is a physically allowed excited state of the system and the PPSs represent the ground state. (b) The order parameter, $|\alpha|^2$, as a function of the detuning $\Delta/U$. (c) The excitation spectrum, $\omega$, on top of the NP (green) and PPS (blue), as a function of the detuning $\Delta/U$ along the orange-dashed cut line in (a). Real (solid) and imaginary (dashed) values with dark (light) hues encoding the particle (hole) excitations, i.e., $ds2>0$ [$ds2<0$]. At the boundary between the white and blue regions in (a), i.e., at $\Delta_{c1}$, $|\alpha|^2$ acquires a finite value in (b), signaling the onset of the $\mathds{Z}_2$ symmetry breaking transition. Correspondingly, excitations on top of the NP, $\pm\omega_{\mathcal{K}}^{\textrm{NP}}$ [cf.~Eqs.\eqref{eq:KPO_NP_eigenfrequencies}], in (c) become fully imaginary at $\Delta_{c1}$ (dashed gray lines). The transition to the coexistence region at $\Delta > \Delta_{c2}$ is once more associated with real excitations on top of the NP. The NP excitations exhibit a norm swap from $\Delta<\Delta_{c1}$ to $\Delta>\Delta_{c2}$ [cf.~Eq.~\eqref{eq:KPO_symplectic_norm}].}
    \label{fig:KPO_phase_diag}
\end{figure}

The time-translation symmetry of the Hamiltonian in the non-rotating frame maps to a $\mathds{Z}_2$ symmetry in the rotating frame (associated with $a\rightarrow-a$)~\cite{Heugel_2019b}. 
The symmetry is spontaneously broken as we tune either the pump strength $G$ or the detuning $\Delta$. In Fig.~\ref{fig:KPO_phase_diag}(a), we draw the corresponding mean-field rotating energy potential landscape,
\begin{equation}
\label{eq:KPO_MF_landscape}
\begin{split}
    \bar{H}_{\mathcal{K}} = -\Delta (\alphare^2+\alphaim^2) + \frac{U}{2}(\alphare^2+\alphaim^2)^2 \\ + G_\re(\alphare^2-\alphaim^2)+2G_\im\alphare\alphaim\,,
\end{split}
\end{equation}
obtained by substituting the mean-field ansatz $a =\alphare + i\alphaim$ into Eq.~\eqref{eq:KPO_Hamiltonian}. The ground state is obtained by minimizing the Hamiltonian Eq.~\eqref{eq:KPO_Hamiltonian}, and calculating the order parameter $|\alpha|^2$, see Appendix~\ref{sec:KPO_closed_appendix} for details. Considering the mean-field energy potential landscape, the phase diagram comprises three qualitatively distinct regions in parameter space with I, a paraboloid potential, where the $\mathds{Z}_2$ symmetry is preserved and the NP is the ground state; II, a double-well potential separated by a saddle, where the $\mathds{Z}_2$ symmetry will be spontaneously broken and either of the PPSs, which are phase shifted by $\pi$, becomes the ground state; and III, a double-well potential separated by a potential hill (maximum), where the PPSs are still the ground states of the system. Crucially, in III the NP is a physical excited state of the system, while in II the NP is unphysical, i.e., its fluctuation spectrum presents complex frequencies, cf.~Sec.~\ref{subsec:closed_sys}. In Fig.~\ref{fig:KPO_phase_diag}(b), we plot the order parameter $|\alpha|^2$ and show that it acquires a finite value within regions II and III, corresponding to the so-called parametric instability~\cite{Nayfeh_book_Nonlinear_Oscillations}.

To corroborate this statement, we follow the procedure outlined in Sec.~\ref{subsec:closed_sys} and find the excitation spectrum associated with the Hamiltonian Eq.~\eqref{eq:KPO_Hamiltonian}. Here, we present the excitations on top of the NP, while the PPS case is detailed in Appendix~\ref{sec:KPO_closed_appendix}. The dynamical matrix associated with the NP reads [cf. Eq.~\eqref{eq:dynamical_matrix_def}]
\begin{align}
	\label{eq:KPO_NP_Hamiltonian}
	H_{\mathcal{K}}^{\textrm{NP}} & = \frac{1}{2}\begin{pmatrix}
					a^\dagger   &   a
				\end{pmatrix}\begin{pmatrix}
					-\Delta & G \\ G^* & -\Delta
				\end{pmatrix}\begin{pmatrix}
					a   \\  a^\dagger
				\end{pmatrix} + \frac{\Delta}{2}\,,\\
	D_{\mathcal{K}}^{\textrm{NP}} & = \frac{1}{2}\begin{pmatrix}
					-\Delta & G \\ -G^* & \Delta
				\end{pmatrix}\,.
\end{align}
In the NP, the cavity is empty and therefore there is no contribution from the Kerr nonlinearity. The eigenvalues of the dynamical matrix describing the excitations are
\begin{equation}
	\label{eq:KPO_NP_eigenfrequencies}
	\omega_{\mathcal{K}}^{\textrm{NP}} = \pm \frac{1}{2}\sqrt{\Delta ^2 - |G|^2}\,.
\end{equation}
In Fig.~\ref{fig:KPO_phase_diag}(c), we plot the excitation spectra on top of the NP and PPS, see also Appendix~\ref{sec:KPO_closed_appendix} for additional details. The NP spectrum becomes fully imaginary along the dashed line cut in Fig.~\ref{fig:KPO_phase_diag}(a) when entering region II (at $\Delta_{c1}$), and at the same time the PPS spectrum becomes real, signaling the onset of the PT. Increasing the detuning further, a transition to region III occurs (at $\Delta_{c2}$), and the NP excitation spectrum becomes real once more. This occurs in concomitance with a change in the sign of the symplectic norm of the excitations,
\begin{equation}
	\label{eq:KPO_symplectic_norm}
	ds_{\mathcal{K}}^2 = \frac{|G|^2}{(\omega_{\mathcal{K}}^{\textrm{NP}} + \Delta)^2}-1\,,
\end{equation}
cf. Eq.~\eqref{eq:symplectic_norm} and see Appendix~\ref{sec:KPO_closed_appendix} for the derivation of Eq.~\eqref{eq:KPO_symplectic_norm}. A positive symplectic norm is a signature of particle-dominated fluctuations, whereas a negative one corresponds to hole-like fluctuations. We therefore observe a swap in the norm of the two NP excitations between regions I and III of the phase diagram, i.e., the white and light-blue regions in Fig.~\ref{fig:KPO_phase_diag}(a). The two NPs are dynamically different, reflecting that in region III the NP is an excited state of the system. We dub it an excited normal phase (e-NP). As we show later, in the presence of dissipation, the e-NP becomes a stable steady state of the system. We here note that excitations on top of the PPS retain their respective norms throughout the entire parameter space, i.e., the positive frequency has a positive norm, whereas the negative one has a negative norm.

\subsubsection{Open system}

\begin{figure*}[t!]
        \includegraphics[width=\textwidth]{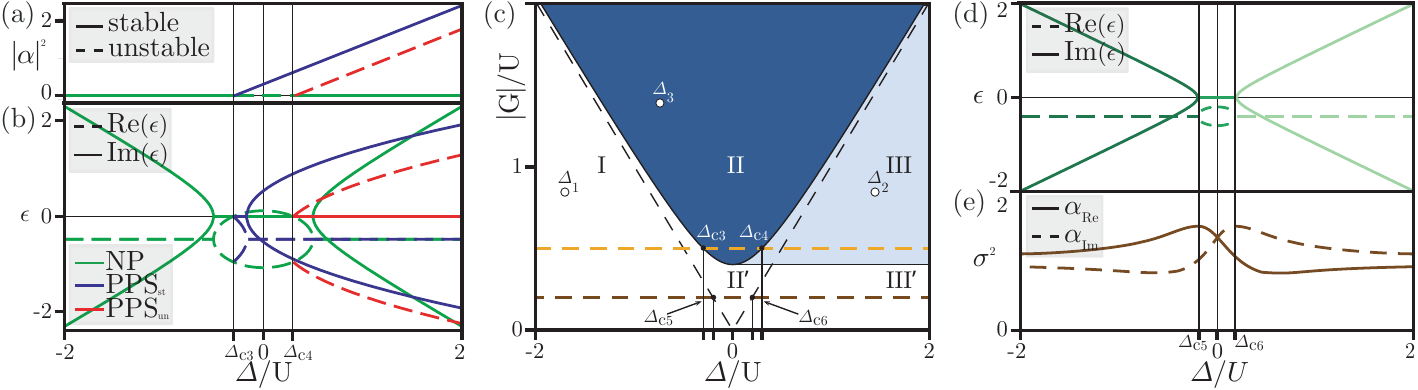}
    \caption{Open Kerr parametric oscillator [cf.~Eq.~\eqref{eq:KPO_master_eq}]. (a) Mean-field steady-state cavity occupation along the orange cut line in (c), solid (dashed) lines indicate (un-)stable solutions. Above $\Delta_{c3}$, the cavity transitions from the normal phase to a high-amplitude solution. For $\Delta > \Delta_{c4}$, a new unstable high-amplitude solution appears and the normal phase becomes stable once more. (b) Real (dashed) and imaginary parts (solid) of the fluctuations (Liouvillian) eigenvalues for the NP (green) and high-amplitude PPS solutions (red, blue). (c) Steady-state phase diagram for $\kappa/U= 0.4$. The normal phase is now the only stable steady state in regions II' and III'. Additionally, the light-blue region is now a region of costable solutions. (d) Eigenvalues behavior and (e) variance for the real (solid) and imaginary (dashed) quadrature of the cavity, on top of the normal phase and along the brown cut line in (c). The fluctuations exhibit a discontinuous exceptional pointlike scenario, where they become overdamped in region II'. The variance does not show any divergent feature in concomitance with the discontinuous fluctuations.}
    \label{fig:KPO_open}
\end{figure*}

Following the methodology introduced in Sec.~\ref{subsec:open_and_steady_states}, we now study the impact of dissipation on the KPO Eq.~\eqref{eq:KPO_Hamiltonian}. We introduce single-photon losses in the form of the same Lindblad dissipator as in Eq.~\eqref{eq:Lindblad_kappa}, and study the master equation:
\begin{equation}
	\label{eq:KPO_master_eq}
	\frac{d \rho}{dt} = -\frac{i}{\hbar}[H_{\mathcal{K}},\rho] + \mathcal{L}[a]\rho\,.
\end{equation}
We proceed and perform a mean-field approximation and characterize the steady state of the system via the expectation value of the operator $\hat{a}$, i.e., via $\alpha = \langle a \rangle$. The equation of motion describing the evolution of $\alpha$ is given by
\begin{equation}
	\label{eq:KPO_EOM_a}
	\frac{d \alpha}{dt} = i(\Delta{\alpha} - U \alpha^* \alpha\,\alpha - G \alpha^*) - \kappa \alpha\,.
\end{equation}
We look for the steady state $d \alpha/dt = 0$, and find the solutions for the order parameter $\alpha$, see Appendix~\ref{sec:KPO_open_appendix} for the analytic expressions of the solutions. In Fig.~\ref{fig:KPO_open}(a), we plot the order parameter behavior as a function of detuning $\Delta$ along the orange cut in Fig.~\ref{fig:KPO_open}(c). From negative to positive detuning, we start from a parameter regime where only one stable solution is possible, i.e., the trivial NP. We then enter regions where up to five solutions for the steady-state mean-field equations are possible. 
Due to the $\mathds{Z}_2$ symmetry of the problem, the non-zero amplitude solutions have pairwise the same amplitude, but are $\pi$-shifted in phase. We therefore focus only on three representative solutions. 

As before, we perform a stability analysis against linear fluctuations and study the eigenvalues of the possible solutions, cf. Sec.~\ref{subsec:open_and_steady_states} and Appendix~\ref{sec:KPO_open_appendix}. As mentioned at the end of Sec.~\ref{subsec:open_and_steady_states}, a solution is unstable if at least one of the fluctuation eigenvalues has a positive real part. In Fig.~\ref{fig:KPO_open}(b), we plot the fluctuation eigenvalues associated with the various physically relevant steady-states. We observe exceptional pointlike scenarios both for the NPs and PPSs. The NP becomes unstable at $\Delta_{c3}$ where a pitchfork bifurcation takes place and stable PPSs appear, denoted PPS\textsubscript{st}. Increasing the detuning further, we reach the point $\Delta_{c4}$ where the low-population solution becomes stable again (e-NP) alongside additional high-density unstable solutions that appear, denoted PPS\textsubscript{un}. The former high-density solutions stay stable, and we have a region of co-stability between the e-NP and the PPS\textsubscript{st}.

We extend our study to the whole parameter space and draw the open steady-state phase diagram in Fig.~\ref{fig:KPO_open}(c). Similar to the closed system case, the phase diagram comprises three different regions: (I) the white region presents only the NP as a stable steady-state, (II) the blue region only the PPS\textsubscript{st}, and, finally, (III) the light-blue one is a region of co-stability between the e-NP and the PPS\textsubscript{st}. At the same time, the open system phase diagram is strikingly different with respect to the closed system one. Specifically, dissipation overwhelms the PPS tendency of the system, and shifts regions II and III upward toward stronger drives $G$, see Fig.~\ref{fig:KPO_phase_diag}(a). In turn, new regions appear: In II', the NP is stabilized despite the fact that it does not correspond to a physical state of the closed system and, in III', the NP remains the sole stable solution while being an excited state of the system.
The appearance of region II' is what we dub a dissipation-induced PT, which in turn directly connects two distinct phases (regions I and III) in the closed system. As we shall see in the following, as the dissipation stabilizes the e-NP in a region of the parameter space where it is an excited state of the closed system, it is a dynamically different NP with respect to the NP at negative detuning.

We conclude this subsection by showing that the NP$\to$ e-NP transition, involving exceptional point-like discontinuities in the fluctuation spectrum, is not a third-order PT. Specifically, we show that the variance of the fluctuations [cf.~Eq.~\eqref{eq:variance_gen}] are differentiable. Without loss of generality, we consider the brown cut in Fig.~\ref{fig:KPO_open}(c), and study the fluctuation variance over the NPs. The results are shown in Figs.~\ref{fig:KPO_open}(d)-\ref{fig:KPO_open}(e). We identify the transition region, $\Delta_{c5}<\Delta<\Delta_{c6}$, as a region of overdamped excitations [cf. Sec.~\ref{subsec:static_HO} and Fig.~\ref{fig:HO_var_eigen_resp}(a)], where the real parts of the eigenvalues, $\rm{Re}(\epsilon)$, split but remain negative and the variances remain finite while changing continuously. The variance is continuous despite the discontinuity in the eigenvalues due to a compensation coming from their associated eigenvectors, see Appendix~\ref{sec:KPO_open_appendix} for details. Note that there is a crucial difference between the overdamped oscillator in Fig.~\ref{fig:HO_var_eigen_resp}(a) and the KPO in Fig.~\ref{fig:KPO_open}. In the former case, the overdamped region appears as a consequence of strong dissipation, whereas in the latter it appears due to the interplay between a weak dissipation channel [Eq.\eqref{eq:Lindblad_kappa}] and the effect of the parametric drive [Eq.\eqref{eq:KPO_Hamiltonian}], leading to a locking of the motion to the drive, which manifests as overdamped fluctuations. 

\subsubsection{Keldysh KPO} 
\label{sub:keldysh_kpo}

We now employ the formalism introduced in Sec.~\ref{subsec:dynamical_observables} and show that the two NPs, that are now connected via the dissipation-stabilized region II', are dynamically different. We first write the Keldysh action Eq.~\eqref{eq:rotated_Keldysh_action} for the KPO,
\begin{align}
    	S = \int dt [& a_c^* i \partial_t a_q + a_q^* i \partial_t a_c + \Delta a_c^*a_q + \Delta a_q^*a_c \notag\\
		&- \frac{U}{2}(|a_c|^2 + |a_q|^2) (a_c^* a_q + a_q^* a_c) \notag\\
		&- G a_c^*a_q^* - G^* a_c a_q \notag\\
		&- i\kappa(a_c^* a_q - a_c a_q^* - 2a_q^*a_q)]\,,
	\label{eq:KPO_action}
\end{align}
where we did not resort to the matrix notation due to the presence of the quartic interaction term. As is customary in the Keldysh path integral formulation~\cite{Kamenev_book,Sieberer_2016}, we perform a saddle-point approximation minimizing the action with respect to the quantum field:
\begin{equation}
    \begin{split}
    \frac{\delta S}{\delta a_q^*} = i\partial_t a_c + \Delta a_c - \frac{U}{2}(a_c^* a_q + a_q^* a_c) a_q \\
    - \frac{U}{2}(|a_c|^2 + |a_q|^2) a_c - G a_c^* + i\kappa (a_c + a_q)\,.    
    \end{split}
\end{equation}
We, then, set $a_c(t) = \sqrt{2}\alpha, a_q = 0$, and solve for the steady state, i.e., $\partial_t a_c=0$. We note here that such a solution is the same on the two branches of the Keldysh contour and coincides with the mean-field solution of Eq.~\eqref{eq:KPO_EOM_a} for the cavity field.

We now study the fluctuations on top of the mean-field solution and address the spectral properties of our system. We do so by expanding around the mean-field solutions
\begin{equation}
\label{eq:KPO_fluc_field}
	a_{c,q} = \alpha + \delta a_{c,q}\,,
\end{equation}
where $\alpha$ is either one of the NPs or PPSs and $\delta a_{cq}$ are the fluctuation fields. We insert Eq.~\eqref{eq:KPO_fluc_field} into the action Eq.~\eqref{eq:KPO_action} and retain up to second order in the fluctuation fields to obtain the fluctuation action
\begin{equation}
\label{eq:KPO_S_fluc}
	S_{\textrm{fluc}} = \int_\omega \frac{1}{2} \begin{pmatrix}\phi_c^\dagger\,\phi_q^\dagger\end{pmatrix} \begin{pmatrix}	0 &	[G^A]^{-1}	\\	[G^R]^{-1}	&	D^K	\end{pmatrix} \begin{pmatrix}\phi_c \\ \phi_q\end{pmatrix}\,,
\end{equation}
where we introduced the spinor notation $\phi_{cl,q} = (\delta a_{cl,q}, \delta a_{cl,q}^*)^T$ and defined the inverse Green's functions and Keldysh component as
\begin{align}
\label{eq:KPO_inv_GR}
	[&G^R]^{-1}(\omega)   
	\\ &= \begin{pmatrix}
	\omega + \Delta - 2U|\alpha|^2 + i\kappa	&	-U \alpha\alpha - G	\\
	-U \alpha^*\alpha^* - G^*	&	-\omega + \Delta - 2U|\alpha|^2 - i\kappa				 \end{pmatrix}\,, \notag
	\end{align}
	\begin{align}
	D^K(\omega) =& 2i\kappa\mathds{I}_{2\times2}\label{eq:KPO_inv_GK}\,, 
\end{align}
with $\mathds{I}_{2\times2}$ the identity matrix of dimension $2$. We now study the spectral properties of the NPs and of the PPSs.

\subsubsection{Normal phase} 
\label{ssub:KPO_normal_phase}

In the NPs, we have a zero mean-field solution, $\alpha=0$. We invert Eq.~\eqref{eq:KPO_S_fluc} with Eqs.~\eqref{eq:KPO_inv_GR} and~\eqref{eq:KPO_inv_GK} to obtain the Green's functions
\begin{widetext}
\begin{align}
	G_{\mathcal{K}}^R(\omega) & = {G_{\mathcal{K}}^A}^\dagger(\omega) = \frac{1}{(\omega + i\kappa)^2 - \Delta^2 + |G|^2}\begin{pmatrix}
	\omega - \Delta + i\kappa	&	- G	\\
	- G^*	&	-\omega - \Delta - i\kappa
													 \end{pmatrix}\,, \notag \\
	G_{\mathcal{K}}^K(\omega) & = \frac{2i\kappa}{((\omega+i\kappa)^2-\Delta^2+|G|^2)((\omega-i\kappa)^2-\Delta^2+|G|^2)} \begin{pmatrix}
	(\omega-\Delta)^2+|G|^2+\kappa^2	&	2G(\Delta - i\kappa)	\\
	- 2G^*(\Delta - i\kappa)	&	(\omega+\Delta)^2+|G|^2+\kappa^2
													 \end{pmatrix}\,. 
													 \label{eq:dynResKPO}
\end{align}
\end{widetext}
We consider two distinct points in the parameter space, $\Delta_1$, and $\Delta_2$ [cf.~Fig.~\ref{fig:KPO_open}(c)], where the NPs are stable steady states. The former lies in the region of negative detuning, where the NP coincides with the ground state of the closed system. The second one instead is in the positive detuning region, where dissipation stabilizes the NP resulting in the e-NP.

\begin{figure}[ht!]
    \includegraphics[width=\columnwidth]{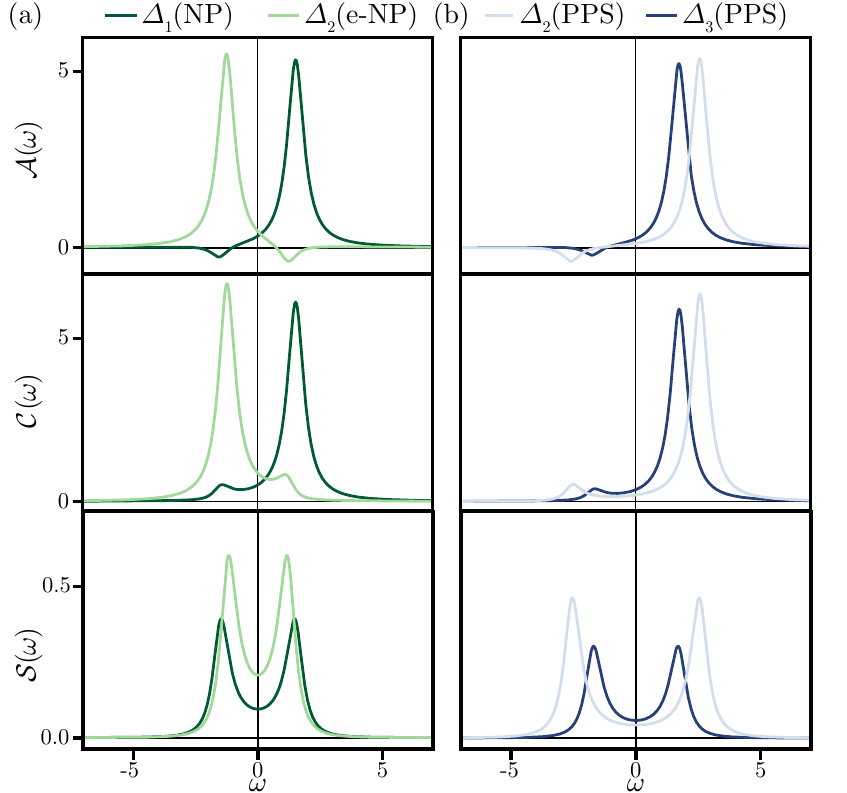}
    \caption{Dynamical responses of the KPO [cf.~Eqs.~\eqref{eq:dynResKPO}]. Top: The spectral function, $\mathcal{A}(\omega)$. Middle: The power spectrum, $\mathcal{C}(\omega)$. Bottom: The fluorescence spectrum, $\mathcal{S}(\omega)$. (a) The responses on top of the NPs, at point $\Delta_1, \Delta_2$ in Fig.~\ref{fig:KPO_open}(c). (b) The responses on top of the PPSs, at points $\Delta_2, \Delta_3$ in Fig.~\ref{fig:KPO_open}(c). The spectral response and power spectrum of the NPs feature a mode inversion when going from the NP to the e-NP,  $\Delta_1\to \Delta_2$. This is a signature of the dynamically different nature of the two normal phases in the different regions of parameter space. On the contrary, for the PPSs, no qualitative change in the spectra is observed--neither when crossing from negative to positive detuning nor when considering the co-stability region.}
    \label{fig:KPO_spectra}
\end{figure}

We calculate the spectral function Eq.~\eqref{eq:spectral_function}, the power spectrum Eq.~\eqref{eq:power_spectrum}, and the fluorescence spectrum Eq.~\eqref{eq:fluorescence_spectrum}. We find that the first two observables present a peak inversion between the two different regions in the parameter space whereas the last is perfectly symmetric, see Fig.~\ref{fig:KPO_spectra}(a). The peak swap between the two region bears important physical significance and underlines the fundamental dynamical difference between the NP and the e-NP. It is reminiscent of the particle-hole inversion in the excitation spectrum of the NP already highlighted in the closed system. More specifically, it also marks the change in oscillation chirality relative to the reference frame [cf. Fig.~\ref{fig:HO_H+L}], by the fact that a negative peak appears at positive response frequency in the spectral function, $\mathcal{A}(\omega)$~\cite{Soriente_2020}. Ultimately, dissipation stabilizes the NP in region II$'$, where it is not a physical solution of the closed system, and therefore does not exhibit proper excitations, as the excitations are overdamped. In region III$'$, an otherwise-inaccessible excited state is reachable by adiabatic ground state evolution from negative to positive detuning. At the same time, the system response retains features of an excited state, signaled by the hole-like nature of the dynamical fluctuations (peak inversion).

\subsubsection{Parametric phase state}
\label{ssub:KPO_parametron}

The PPSs are characterized by a finite cavity field $\alpha$, see Appendix~\ref{sec:KPO_open_appendix}. We invert Eqs.~\eqref{eq:KPO_inv_GR} and \eqref{eq:KPO_inv_GK} once more with the new mean-field and obtain the Green's functions for the PPSs. As for the NPs, we consider two points in the parameter space, $\Delta_2$, and $\Delta_3$ in Fig.~\ref{fig:KPO_open}(c). The point $\Delta_3$ lies in region II, where only the PPS\textsubscript{st} are stable steady states, whereas $\Delta_2$ is positioned in the coexistence region III. In Fig.~\ref{fig:KPO_spectra}(b), we plot the dynamical response functions on top of the PPS\textsubscript{st} at $\Delta_2$, and $\Delta_3$. There are no qualitative changes in the spectra as opposed to the NPs, which undergo a peak swap between the two regions in parameter space. This agrees with the discussion above on the differences that the closed system experiences between the NP and e-NP, whereas the PPSs do not present a particle-hole fluctuation inversion.

\subsection{Interpolating Dicke-Tavis-Cummings}
\label{sec:IDTC}

In the previous section, we studied the KPO as a ubiquitous model describing a broad class of systems, where one directly drives a bosonic resonator (be it vibrational, electric, photonic, or composed of more complex particles) with a two-particle drive. In this section, we consider instead a many-body light-matter system, where PTs appear due to the coupling between the matter degrees of freedom and a bosonic resonator cavity~\cite{Dicke_1954,Soriente_2018,Dogra_2019,Zhao_2019,Chiacchio_2019,Puebla_2019,Zupancic_2019,Zhu_2020,Zhang_2020,Zhu_2020_PRR}. We specifically study a ubiquitous model describing light-matter systems, the IDTC model~\cite{Baksic_2014, Soriente_2020,Bhaseen_2012,Stitely_2020_PRR}, which is a generalized version of the Dicke~\cite{Dicke_1954,Dimer_2007,Baumann_2010} and the Tavis-Cummings~\cite{Tavis_1968} models. The model system comprises of a single-mode bosonic cavity (light) coupled to the $x$ and $y$ components of $N$ spinlike (two levels) degrees of freedom (matter). As a function of the coupling between the cavity and the spins, the IDTC model features transitions between an empty cavity state, the NP, to a highly occupied cavity state, dubbed superradiant phase (SP)~\cite{Zhiqiang_2017,Kirton_2019}.

In similitude to the KPO above, dissipation stabilizes and extends the NP into a new parameter regime, dubbed e-NP~\cite{Soriente_2018, Soriente_2020}. This dissipation-induced PT also allows for regions of coexistence between the e-NP and SP~\cite{Soriente_2018,Stitely_2020}. The analogy between the KPO and IDTC is even more explicit when considering the fluctuation spectrum, which shows a soft-mode invertion along the low-density dissipation-facilitated NP $\rightarrow$ e-NP transition. We highlight here these distinctive features of the dissipative IDTC, and draw the universal parallels between this many-body light-matter system and the KPO, while employing the methodology introduced in the previous sections.

\subsubsection{Closed system}

\begin{figure}[t!]
    \includegraphics[width=\columnwidth]{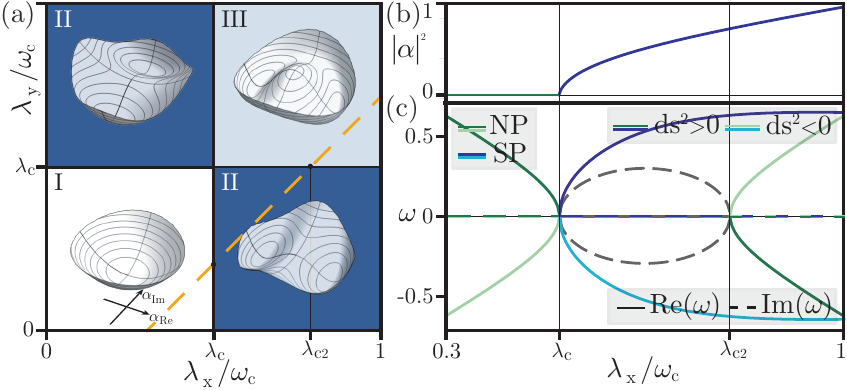}
    \caption{Closed interpolating Dicke-Tavis-Cummings model [cf.~Eq.~\eqref{eq:H_IDTC}]. (a) Phase diagram: Similar to the KPO (cf.~Fig.~\ref{fig:KPO_phase_diag}), the system has $\mathds{Z}_2$ symmetries and is characterized by three different regions; with $\omega_c=\omega_z$ and schematic illustration of the mean-field energy potential as a function of the real and imaginary parts of the cavity field, $\alphare$ and $\alphaim$, respectively. Quantum phase transitions occur between a normal phase (white) and a symmetry broken phase (blue). Analogous to the KPO, the light-blue region indicates a parameter regime where the normal phase is an excited state of the system and the SPs represent the ground state. (b) The order parameter, $|\alpha|^2$, and (c) the excitation spectrum, $\omega$, on top of the NP (SPs), as a function of the coupling along the orange-dashed cut line in (a). In (c), real (solid) and imaginary (dashed) parts on top of the NP (green) and SPs green (blue), dark (light) hues encode the particle (hole) excitations, i.e., $ds2>0$ [$ds2<0$]. The excitation behavior is analogous to the KPO. At the boundary, $\lambda_c$, between the white and blue region in (a), $|\alpha|^2$ acquires a finite value in (b). Concomitantly, the eigenfrequencies on top of the NP $\omega_{\mathcal{I}}^{NP}$ [cf. Eqs.~\eqref{eq:excitations_IDTC}] in (c) become fully imaginary (dashed gray lines), thus signaling the onset of the symmetry-breaking transition. When $\lambda>\lambda_{c1}$ the eigenvalues return to be real-valued, marking the transition to the coexisting region. Mirroring the KPO, the NP excitations exhibit a norm swap from $\lambda<\lambda_{c}$ to $\lambda>\lambda_{c1}$.}
    \label{fig:IDTC_closed}
\end{figure}

We start our discussion with the closed system IDTC Hamiltonian~\cite{Baksic_2014, Fan_2014, Soriente_2018, Soriente_2020}
\begin{multline}
	\label{eq:H_IDTC}
	H_{\mathcal{I}} = \hbar \omega_c a^\dagger a + \hbar \omega_z S_z \\
	+ \frac{2\hbar\lambda_x}{\sqrt{N}}S_x(a + a^\dagger) + \frac{2\hbar\lambda_y}{\sqrt{N}}iS_y(a - a^\dagger)\,,
\end{multline}
where $a$ is the bosonic annihilation operator of the cavity field, and $\omega_c$ is the cavity's resonance frequency. The ensemble of $N$ two-level system is described in terms of collective spin operators $S_{x,y,z}=\sum_{i=1}^{N}\sigma_{x,y,z}^{\{i\}}$, with Pauli matrices $\sigma_{j}^{\{i\}}$ describing the $i$th two-level system. Due to the tunable coupling between the spins and the cavity, this model interpolates between the Tavis-Cummings ($\lambda_x=\lambda_y$)~\cite{Tavis_1968} and Dicke ($\lambda_x\ne\lambda_y$)~\cite{Dicke_1954,Dimer_2007,Baumann_2010} models.

The IDTC Hamiltonian Eq.~\eqref{eq:H_IDTC} has a $\mathds{Z}_2\times \mathds{Z}_2$ symmetry corresponding to four different superradiant states with finite spin magnetization either along the $x$ or $y$ directions. Furthermore, tuning the couplings to be equal, $\lambda_x=\lambda_y$, boosts the symmetry to become $U(1)$. In either case, increasing the coupling above a critical coupling $\lambda_{ c}=\sqrt{\omega_c\omega_z}/2$ leads to spontaneous breaking of the symmetries, into a SP with finite in-plane magnetization~\cite{Oztop_2012,Kirton_2019}. This is shown in the phase diagram in Fig.~\ref{fig:IDTC_closed}(a). 
The mean-field energy potential landscape lies over the phase space of the resonator and the Bloch sphere of the collective spin. Focusing on the south pole of the Bloch sphere, where the NP lies, it suffices to draw the impact of the spins on the mean-field energy functional of the resonator, see Fig.~\ref{fig:IDTC_closed}(a). The phase diagram comprises of three qualitatively distinct regions in parameter space: I, a paraboloid potential, where the $\mathds{Z}_2\times \mathds{Z}_2/U(1)$ symmetry is preserved and the NP is the ground state; II, a double-well potential separated by a saddle, where the $\mathds{Z}_2\times \mathds{Z}_2$ symmetry is spontaneously broken either along $x$ or $y$ directions, and either of the positive or negative magnetization state becomes the ground state; and III, a double-well potential separated by a potential hill (maximum), where the SPs are still the ground states of the system. Crucially, as in the KPO, in III, the NP is a physical excited state of the system, while in II, the NP is unphysical (imaginary excitation frequencies).

Following the procedure outlined in Sec.~\ref{subsec:closed_sys}, we find the ground state, order parameters $\alpha,X,Y$, and spectrum of excitations of the IDTC model, Eq.~\eqref{eq:H_IDTC}. In the top panel of Fig.~\ref{fig:IDTC_closed}(b), we plot the order parameters $|\alpha|^2$ and show that it acquires a finite value within regions II and III, corresponding to the superradiant PTs, this parallels the NP$\to$PPS of the KPO, cf.~Fig.~\ref{fig:KPO_phase_diag}. To show the existence of the e-NP phase and draw a comparison with the KPO studied above, see Sec.~\ref{subsec:KPO_closed}, we consider the IDTC's normal phase and analyze its excitation spectrum. Details regarding the SP can be found in Ref.~\cite{Soriente_2020}.
The NP fluctuation Hamiltonian and associated dynamical matrix for the IDTC model are
\begin{align}
	\label{eq:IDYC_NP_Hamiltonian}
	H_{\mathcal{I}}^{\textrm{NP}} & = \begin{pmatrix}
					\omega_c    &   \lambda_x+\lambda_y  &   0  &   \lambda_x-\lambda_y \\
					\lambda_x+\lambda_y &   \omega_z & \lambda_x-\lambda_y  &   0 \\
					0   &   \lambda_x-\lambda_y &   \omega_c    &   \lambda_x+\lambda_y \\
					\lambda_x-\lambda_y &   0   &   \lambda_x+\lambda_y &   \omega_z
				\end{pmatrix}\,, \\
	D_{\mathcal{I}}^{\textrm{NP}} & = \begin{pmatrix}
					\omega_c    &   \lambda_x+\lambda_y  &   0  &   \lambda_x-\lambda_y \\
					\lambda_x+\lambda_y &   \omega_z & \lambda_x-\lambda_y  &   0 \\
					0   &   -\lambda_x+\lambda_y &   -\omega_c    &   -\lambda_x-\lambda_y \\
					-\lambda_x+\lambda_y &   0   &   -\lambda_x-\lambda_y &   -\omega_z
				\end{pmatrix}\,,
\end{align}
therefore, we find the closed system eigenvalues~\cite{Soriente_2020}, cf. Sec.~\ref{subsec:closed_sys},
\begin{align}
\label{eq:excitations_IDTC}
    & \left({\omega_{\mathcal{I}}^{\textrm{NP}}}\right)^2= \frac{1}{2}(4(\lambda_x^2+\lambda_y^2)+\omega_c^2+\omega_z^2\\
    & \pm\sqrt{16(\lambda_x^2-\lambda_y^2)^2+8(\lambda_x^2+\lambda_y^2)(\omega_c+\omega_z)^2+(\omega_c^2-\omega_z^2)^2})\,. \notag
\end{align}
Unlike the KPO, the IDTC exhibits two excitation eigenmodes. We focus on the soft-mode [with the negative sign in Eq.~\eqref{eq:excitations_IDTC}, plotted in Fig.~\ref{fig:IDTC_closed}(c)] since it exhibits an interesting behavior. In close analogy with the KPO, the NP soft-mode excitation eigenfrequencies become imaginary upon crossing of the superradiant boundary, $\lambda_c$. They, then, become real again when crossing to the light-blue region III at $\lambda_{c2}$. As discussed in Ref.~\cite{Soriente_2020}, the transition from region I to III occurs alongside with a norm flip [cf. Eq.~\eqref{eq:symplectic_norm}] in the soft excitation eigenmode. This flip signals a particle-hole inversion in the excitation spectrum describing the system, marking the e-NP as an excited state of the system. As in the KPO, this feature survives the introduction of dissipation and, in the next section, we show that the e-NP becomes a dissipation-stabilized steady state. The SP excitation spectrum does not show any norm sign changes~\cite{Soriente_2020}.

\subsubsection{Open system}

\begin{figure*}[t!]
    \includegraphics[width=\textwidth]{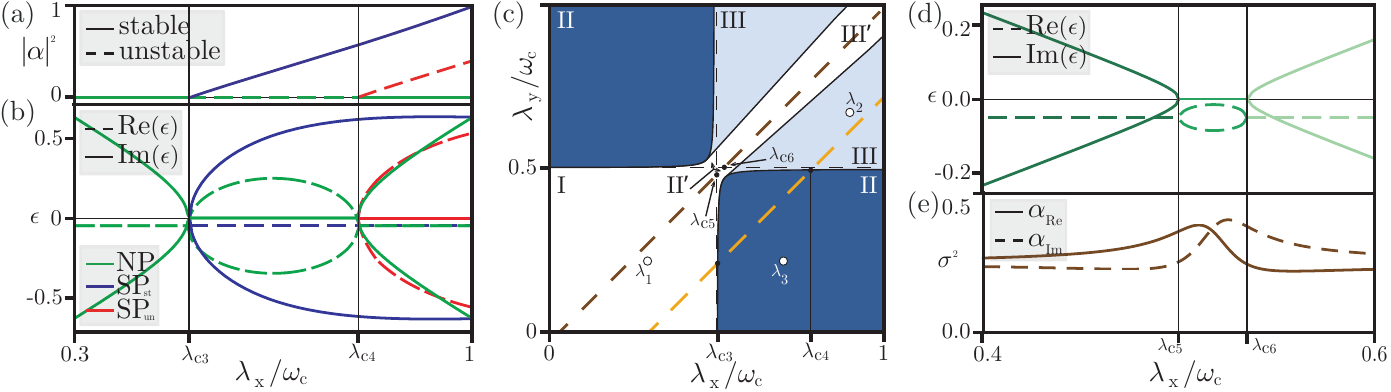}
    \caption{Open IDTC model [cf.~Eq.~\eqref{eq:IDTC_master_eq}]. (a) Mean-field steady-state cavity occupation along the orange cut line in (c), solid (dashed) lines indicate (un-)stable solutions. Above $\lambda_{c2}$ the cavity transitions from the empty normal phase to a high-density (superradiant) solution. For $\lambda > \lambda_{c3}$, a new unstable high-density solution appears and the normal phase becomes stable again. (b) Real (dashed) and imaginary parts (solid) of the fluctuations (Liouvillian) eigenvalues for the NP (green) and SP solutions (red, blue). (c) Steady-state phase diagram of the dissipative model [Eq.~\eqref{eq:IDTC_master_eq}] for $\kappa/\omega_c= 0.1$. The normal phase is now the only stable steady state in the dissipation-induced regions II' and III', where the NP is not the ground state of the closed system [cf.~Fig.~\ref{fig:IDTC_closed}]. Additionally, the light-blue region III becomes a region of ‘‘co-stable" solutions. Adapted from Ref.~\cite{Soriente_2018}. (d) Fluctuation spectrum and (e) variance for the real (solid) and imaginary (dashed) quadrature of the cavity on top of the NPs, along the brown cut line in (c). The variance does not show any divergent feature in concomitance with the appearance of the exceptional point region.}
    \label{fig:IDTC_open}
\end{figure*}

In a previous work~\cite{Soriente_2018}, we analyzed the impact of adding single-photon losses to the Hamiltonian~\eqref{eq:H_IDTC}, cf.~Eq.~\eqref{eq:Lindblad_kappa}. The dissipator drastically changes the phase diagram. We here reiterate, mirroring Sec.~\ref{subsec:open_and_steady_states}, the main steps that led to the modified phase diagram to stress the close similarities between the KPO and the IDTC.

We start from the master equation,
\begin{equation}
	\label{eq:IDTC_master_eq}
	\frac{d \rho}{dt} = -\frac{i}{\hbar}[H_{\mathcal{I}},\rho] + \mathcal{L}[a]\rho\,,
\end{equation}
where $\mathcal{L}[a]\rho$. We define the mean-field order parameters according to $\langle a\rangle = \sqrt{N}\alpha, \langle S_i\rangle = N i$ with $i = X,Y,Z$, and study the associated equations of motion [cf. Eq.\eqref{eq:EOM_matrix}]:
\begin{equation}
    \begin{aligned}
    	\label{eq:IDTC_MF_EOM}
    	\frac{d\alpha}{dt} & = - i \omega_0 \alpha - i \frac{2\lambda_x}{\sqrt{N}}X - \frac{2\lambda_y}{\sqrt{N}}Y - \kappa \alpha\,, \\
    	\frac{dX}{dt} & = - \omega_z Y - i\frac{2\lambda_y}{\sqrt{N}}(\alpha^* - \alpha) Z \,,\\
    	\frac{dY}{dt} & = \omega_z X - \frac{2\lambda_x}{\sqrt{N}}(\alpha^* + \alpha) Z \,,\\
    	\frac{dZ}{dt} & = \frac{2\lambda_x}{\sqrt{N}}(\alpha^* + \alpha) Y + i\frac{2\lambda_y}{\sqrt{N}}(\alpha^* - \alpha) X\,.
    \end{aligned}
\end{equation}
It is possible to find an analytical expression for the steady-state solutions~\cite{Soriente_2018}, and in Fig.~\ref{fig:IDTC_open}(a), we plot the resulting stationary order parameter behavior along the orange cut in Fig.~\ref{fig:IDTC_open}(c). As for the open KPO [cf.~Fig.~\ref{fig:KPO_open}], initially only the trivial NP solution is a stable steady-state. Then, in regions II and III, up to four additional non-trivial superradiant solutions are allowed. Given the $\mathds{Z}_2$ symmetries of the problem, the non-zero solutions appear pairwise opposite of each other in phase space. 
We study the stability of the solutions against linear fluctuations [cf. Eq.\eqref{eq:fluc_stability}] and identify the stable and unstable solutions. In Fig.~\ref{fig:IDTC_open}(b), we plot the eigenvalues associated with the possible steady-states. The NP becomes unstable at $\lambda_{c3}$, where the superradiant transition takes place. Increasing the coupling further, the co-stability region III appears, where the low-population solution (e-NP) becomes stable again at $\lambda_{c4}$, and an unstable SP appears, SP\textsubscript{un}. Stable (unstable) solutions are highlighted as solid (dashed) lines in Fig.~\ref{fig:IDTC_open}(a). 

The open steady-state phase diagram is plotted in Fig.~\ref{fig:IDTC_open}(c). Mapping one-to-one to the KPO, the open IDTC phase diagram includes the variety of regions, I, II, III, II$'$, and III$'$.
Specifically, due to dissipation, the superradiant regions in Fig.~\ref{fig:IDTC_open}(c) are now separated by a NP sliver (II') that connects two regions (I and III'), where the NPs are the only stable steady state. Stressing the KPO $\leftrightarrow$ IDTC analogy further, dissipation stabilizes the NP in region III' of the parameter space, where it is an excited state of the closed system, this justifies the name e-NP~\cite{Soriente_2020}. 

In Figs.~\ref{fig:IDTC_open}(d) and~\ref{fig:IDTC_open}(e), we study the eigenvalues and variance of the IDTC NPs along a cut line that does not encompass the symmetry broken phase, brown cut in Fig.~\ref{fig:IDTC_open}(c), cf.~Figs.~\ref{fig:KPO_open}(d) and~\ref{fig:KPO_open}(e). The fluctuations' eigenvalues in region II' show the same overdamped behavior as the KPO (cf. Sec.~\ref{subsec:static_HO}, Fig~\ref{fig:KPO_open}) with the imaginary parts coalescing to zero, and the real parts splitting but remaining negative. This is accompanied by a finite variance squeezing along the transition between the NP and the e-NP. Here, too, the exception point-like discontinuity in the fluctuations' spectrum does not lead to sharp transitions in the variance. In conclusion, the open IDTC model exemplifies another (many-body) family of systems, where dissipative effects are enhanced by the interactions among the components of the system, leading to dissipation-induced PTs (region II') and the plethora of associated corollary signatures. In the following, we turn to highlight that similar dynamical signatures appear in this model as well.

\subsubsection{Keldysh IDTC}

We conclude our treatment of the IDTC model using the formalism introduced in Sec.~\ref{subsec:dynamical_observables}, to show that the two NPs are dynamically different. To find a path integral representation for the Hamiltonian Eq.~\eqref{eq:H_IDTC}, we bosonize the spin around the noninteracting ground state. The ground state is characterized by the magnetization $S_z = -N/2$. The bosonized spin is obtained using the Holstein–Primakoff transformation, $S_+ = b^\dagger\big(\sqrt{N - b^\dagger b}\big)$ and $S_z=-\frac{N}{2}+b^\dagger b$, where $b, b^\dagger$ are bosonic operators.
The resulting Keldysh action [cf.~Eq.~\eqref{eq:rotated_frequency_Keldysh_action}] for the thus-transformed IDTC model is
\begin{equation}
	\label{eq:S_IDTC}
	S = \frac{1}{2}\int_\omega \; \Phi^\dagger(\omega)\begin{pmatrix} 0 & {\big[G_{4\times 4}^A\big]}^{-1} \\
													{\big[G_{4\times 4}^R\big]}^{-1} & {D_{4\times 4}^K}^{-1} \end{pmatrix} \Phi(\omega),
\end{equation}
where we introduced the eight-components cavity-spin Nambu spinor, $\Phi = [\Phi_{c}(\omega), \Phi_{q}(\omega)]$, defined as the concatenated classical and quantum four-spinors: $\Phi_i(\omega) = (a_i(\omega), a^*_i (-\omega), b_i(\omega), b^*_i (-\omega))^T$ with $i =c,q$, respectively. For the explicit form of the Green's functions, $G^R_{4\times4},G^A_{4\times4}$ and the Keldysh self-energy, $D^K_{4\times4}$, in Nambu space, see Ref.~\cite{Soriente_2020}.

Varying the action Eq.~\eqref{eq:S_IDTC} with respect to the quantum components of the fields and substituting $a_c(t) = \sqrt{2}\alpha, b_c(t) = \sqrt{2}\beta, a_q = 0, b_q = 0$, we obtain the coupled equations
\begin{align}
		\frac{\partial S}{\partial a_q^*} & = (-\omega_0 + i\kappa)\alpha \\
		& - 2 \sqrt{1 - |\beta|^2}\big(\lambda_x \beta_\re + i \lambda_y \beta_\im\big) = 0\;, \notag \\
		\frac{\partial S}{\partial b_q^*} & =  - \omega_z \beta - 2 i \beta_\re \beta_\im \frac{\lambda_x \alphare - i \lambda_y \alphaim}{\sqrt{1 - |\beta|^2}} \\
		& + 2 \frac{\big(\lambda_x \alphare \beta_\im^2+ i \lambda_y \alphaim \beta_\re^2 - \lambda_x \alphare - i \lambda_y\big)}{\sqrt{1 - |\beta|^2}} = 0\,, \notag
\end{align}
where $\alpha,\beta \in\mathbb{C}$. We can split them into their real and imaginary parts 
and recast them in terms of the mean-field order parameters $\alpha, X, Y, Z$ using the inverse Holstein-Primakoff transformation. This procedure yields the mean-field equations Eqs.~\eqref{eq:IDTC_MF_EOM}.

We turn now to study the fluctuations on top of the mean-field solutions, and address the spectral properties of the open IDTC. The different stationary phases of the IDTC can be described by the field $a$ ($b$) given by the mean-field solutions $\alpha$ ($\beta$), and a fluctuating term, i.e., $a \to \alpha + \delta a$ ($b \to \beta + \delta b$) [cf.~similar treatment of the KPO in Eqs.~\eqref{eq:KPO_fluc_field} and~\eqref{eq:KPO_S_fluc}]. The resulting IDTC quadratic fluctuation action is
\begin{equation}
\label{eq:IDTC_S_fluc}
	S_{\textrm{fluc}} = \frac{1}{2}\int_\omega \; \delta\Phi^\dagger(\omega)\begin{pmatrix} 0 & {\big[G_{4\times 4}^A\big]}^{-1} \\
													{\big[G_{4\times 4}^R\big]}^{-1} & {D_{4\times 4}^K} \end{pmatrix} \delta\Phi(\omega),
\end{equation}
with the Green's functions
\begin{align}
\label{eq:IDTC_inv_GR}
    &{\big[G_{4\times 4}^R\big]}^{-1} \\
    &\resizebox{.88\hsize}{!}{$\displaystyle={\begin{pmatrix} \omega -\omega_0 + i\kappa & 				 0 			   &				-\bar{\lambda}_1^* 					&				 -\bar{\lambda}_2 						\\
	0 				 & -\omega -\omega_0 - i\kappa &				-\bar{\lambda}_2^* 					&				 -\bar{\lambda}_1 						\\
	-\bar{\lambda}_1 		 &			 -\bar{\lambda}_2 	   & \omega - \omega_z - \delta\bar{\omega}_1 	&		 -\delta\bar{\omega}_2^* 					\\
	-\bar{\lambda}_2^* 	 &			 -\bar{\lambda}_1^* 	   & 		 -\delta\bar{\omega}_2 				& -\omega - \omega_z - \delta\bar{\omega}_1^*						\end{pmatrix}\,,}$}
	\notag
\end{align}
and the Keldysh self-energy
\begin{equation}
	D_{4\times 4}^K = 2i\text{diag}(\kappa,\kappa,0,0).
\end{equation}
The definition of the coefficients $\delta\bar{\omega}_1,\delta\bar{\omega}_2,\bar{\lambda}_1,\bar{\lambda}_2$ can be found in Ref.~\cite{Soriente_2020}.

To obtain the photon-only action, we integrate out the Holstein-Primakoff fluctuation field $\delta b$ and obtain a Keldysh functional integral only over the photon fields. The resulting photon-only action reads
\begin{multline}
\label{eq:IDTC_S_photon}
	S_{\textrm{photon}}[\delta a^*,\delta a] \\
	= \int_\omega A_4^\dagger(\omega)\begin{pmatrix}	0	&	\big[G_{2\times 2}^{A,p}\big]^{-1}(\omega)	\\
	\big[G_{2\times 2}^{R,p}\big]^{-1}(\omega) & D_{2\times 2}^{K,p}(\omega) \end{pmatrix} A_4(\omega)\,,
\end{multline}
where $A_4(\omega) = (\delta a_c(\omega), \delta a_c^*(-\omega), \delta a_q(\omega), \delta a_q^*(-\omega))^T$ is the photon four-vector that collects the classical and quantum field components. The inverse retarded Green's function of the photon is
\footnotesize
\begin{multline}
\label{eq:IDTC_inv_GR_p}
	{\big[G_{2\times 2}^{R}\big]}^{-1} \\
	= \begin{pmatrix}	\omega -\omega_0 + i\kappa  + \Sigma_1^{R}(\omega)	&	\Sigma_2^{R}(\omega)	\\
	\big[\Sigma_2^{R}(-\omega)\big]^*	&	-\omega -\omega_0 - i\kappa + \big[\Sigma_1^{R}(-\omega)\big]^* \end{pmatrix}\,,
\end{multline}
\normalsize
while the Keldysh component of the photon action is
\begin{equation}
\label{eq:IDTC_inv_GK}
	D_{2\times 2}^{K,p} = 2i\kappa \mathcal{I}_{2\times2}\,.
\end{equation}
We now study the spectral properties of the NPs and of the SP.

\subsubsection{Normal phase} 
\label{ssub:IDTC_normal_phase}

In the NPs, we have as stationary mean-field solution, $\alpha=0, X=Y=0, Z=-1/2$. Inverting Eq.~\eqref{eq:IDTC_S_photon} and using Eqs.~\eqref{eq:IDTC_inv_GR_p} and~\eqref{eq:IDTC_inv_GK}, we obtain the photon Green's functions. The analytical expressions are quite lengthy, and we report here only the retarded Green's function:
\begin{widetext}
\begin{equation}
\label{eq:dynRespIDCT}
	G_{\mathcal{I}}^R(\omega) = \frac{2 \omega_z (\lambda_x^2 + \lambda_y^2)-4 \lambda_x \lambda_y \omega +(\omega^2 -\omega_z^2) (i \kappa +\omega +\omega_0)}{16 \lambda_x^2\lambda_y^2-8 \lambda_x \lambda_y \omega  (\omega +i\kappa)-4 \omega_0 \omega_z (\lambda_x^2+\lambda_y^2)+(\omega ^2-\omega_z^2) ((\omega+i\kappa)^2-\omega_0^2)}\,.
\end{equation}
\end{widetext}
We focus on the points $\lambda_1$ and $\lambda_2$ marked in Fig.~\ref{fig:IDTC_open}(c), both of which are in the parameter space where the NPs are a stable steady state. The point $\lambda_1$ lies in region I below criticality, where the NP coincides with the ground state of the closed system. The point $\lambda_2$ lies instead in the critical region III$'$, where the e-NP is stabilized by dissipation.

\begin{figure}[ht!]
    \includegraphics[width=\columnwidth]{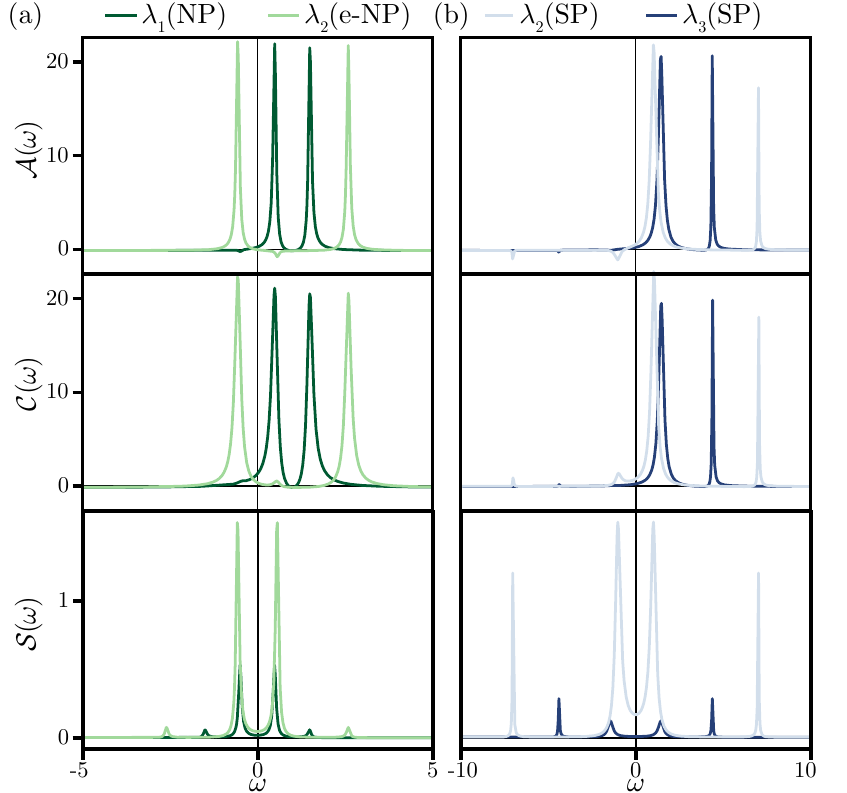}
    \caption{Dynamical responses of the IDTC [cf.~Eq.~\eqref{eq:dynRespIDCT}]. Top: The spectral response, $\mathcal{A}(\omega)$. Middle: The power spectrum, $\mathcal{C}(\omega)$. Bottom: The fluorescence spectrum, $\mathcal{S}(\omega)$, in (a) on top of the NPs, at points $\lambda_1, \lambda_2$, and in (b) on top of the SPs, at points $\lambda_2, \lambda_3$, see Fig.~\ref{fig:IDTC_open}(c). The spectral response and power spectrum of the NPs feature a soft-mode peak inversion when going from the NP to the e-NP, $\lambda_1\to \lambda_2$. This is a signature of the dynamically different nature of the two normal phases in the different regions of parameter space. In comparison, the SPs show no qualitative change in the spectra, neither upon crossing the superradiant threshold nor when considering the co-stability region.}
    \label{fig:IDTC_spectra}
\end{figure}

In Fig.~\ref{fig:IDTC_spectra}(a), we plot the spectral function Eq.~\eqref{eq:spectral_function}, the power spectrum Eq.~\eqref{eq:power_spectrum}, and the fluorescence spectrum Eq.~\eqref{eq:fluorescence_spectrum}. Similar to the KPO case [cf.~Fig.~\ref{fig:KPO_spectra}], we find a soft-mode peak inversion between the two different NP regions. The soft-mode peak swap marks the fact that the e-NP is an excited state of the closed system, which is dynamically distinct from the NP~\cite{Soriente_2020}. This is a final dynamical feature that highlights the universal phenomenology of dissipation-induced PTs, and connects between the KPO and IDTC models. In both systems, coherent interactions facilitate that a weak dissipation channel stabilizes an excited state of the closed system, and signatures of the dissipation-induced stabilization manifest in the dynamical response of the system.

\subsubsection{Superradiant phase} 
\label{ssub:IDTC_superradiant_phase}

In Fig.~\ref{fig:IDTC_spectra}(b), we plot the dynamical responses on top of the superradiant solutions for points $\lambda_2$, and $\lambda_3$ in parameter space, see Fig.~\ref{fig:IDTC_open}(c). The former lies in the coexistence region, whereas the latter lies in the region where only the SP\textsubscript{st} is the stable steady-state. The SP, as the PPS in the KPO, does not present qualitative changes in the spectra, marking once more the close analogy between the two \textit{a priori} distinct physical systems.

\section{Discussion and outlook}
\label{sec:conclusion}

At equilibrium, different phases and transitions between them are well understood using statistical physics arguments, based on equilibrium distribution functions. Thus, a coupling to a thermal bath affects the distribution function, and at $T=0$ (quantum) PTs occur only between ground states of the system. Moving to an out-of-equilibrium setting, the presence of drives and excitations leaking out into the environment, requires us to break away from the framework of equilibrium distribution functions. Crucially, in such open systems, as we also highlight in this paper, a $T=0$ environment does not imply a quantum PT in the resulting phase diagram of the system, i.e., the ground state does not necessarily correspond to the new stationary state of the system coupled to the environment.

Over the years, a plethora of methods have been applied to analyze out-of-equilibrium PTs, including diagonalization of Liouvillians~\cite{Kessler_2012} or non-Hermitian Hamiltonians (cf.~third quantizations~\cite{Prosen_2008}), Keldysh action approach~\cite{Kamenev_book,Sieberer_2016}, and equations of motion~\cite{Petruccione_book} to name a few. Various observables were devised to characterize scenarios appearing in the out of equilibrium, e.g., exceptional points marking stabilized regions due to gain and loss~\cite{Heiss_2012,Ozdemir_2019}, and negative peaks in response functions as signatures of stabilized excited states~\cite{Soriente_2020,Scarlatella_2019}. In this paper, we collected such observations under a unifying framework and related them to simple underlying physical implications, using a mean-field analysis on top of which dynamical fluctuation and responses were explored. 

We thus showed how dissipation can profoundly change the closed system physics, by lifting the boundaries of the closed system phases. We, specifically, highlighted two scenarios that arise when dissipation overwhelms a symmetry-breaking term in the system: both cases involved the stabilization of a phase with higher energy than that of the ground state of the system, either corresponding to an excited physical state or to an unphysical state of the closed system. The latter exhibited an exceptional point-like behavior, where the system is locked to the environment and its fluctuations become overdamped. On the other hand, when dissipation stabilizes a maxima of the closed system Ginzburg-Landau potential (i.e., an excited state), we observed an inversion in the dynamical response function, in accord with the inverted curvature of the potential. Crucially, the appearance of both types of scenarios is correlated to one another, where the overdamped region bridges between areas in phase space where the same state can be a ground or excited state. 

We demonstrated the general phenomenology through two prominent example of open systems, namely, the KPO and the IDTC model. These models mark a wide family of experimental out-of-equilibrium domains, where nonlinearities, dissipation, and time-dependent drives interplay with one another~\cite{Kessler_2012,Leghtas_2015,Chitra_2015,Booker_2020}. Examples thereof range from nonlinear mechanical, electric, and photonic resonators, to driven light-matter systems in a similarly broad range of frequencies, including collective PTs in cold-atom experiments~\cite{Landini_2018,Morales_2019,Chen_2020_SR,Zhu_2020_PRL}, and light-induced modifications of material properties~\cite{Rajasekaran_2016,Schlawin_2019}. As such, we believe that our work paves the way for a systematic study of dissipative stabilization of high-energy states, where key future directions can involve beyond mean-field analysis of the steady state and the introduction of more complex dissipation channels with their own symmetry-breaking tendencies, e.g., cat-state stabilization via two-photon losses and superradiant suppression via spin dephasing~\cite{Leghtas_2015,Kirton_2017}.

\begin{acknowledgments}
We would like to thank A. Eichler, T. Donner, and T. Esslinger for fruitful discussions. We acknowledge financial support from the Swiss National Science Foundation through Grants No. PP00P2\_1163818 and No. PP00P2\_190078, and the ETH Research Grant No. ETH-45 17-1.
\end{acknowledgments}

\appendix

\section{Overdamped harmonic oscillator}
\label{sec:HO_overdamped_appendix}

In this Appendix, we provide additional details on the analysis of the overdamped harmonic oscillator [cf.~Sec.~\ref{subsec:static_HO}], including its fluctuation eigenvalues and variance under noise. Note that the treatment here relies on a stochastic classical approach, which is complementary to that employed for weak dissipation in Sec.~\ref{sec:closed_open_systems}.

The overdamped harmonic oscillator can be studied starting from the classical Newton's equation of motion,
\begin{equation}
    \ddot{x}(t) + \omega_0^2 x(t) + \kappa \dot{x}(t) = \xi(t),
\end{equation}
where $\xi(t)$ is a white noise process with $\langle \xi(t) \xi(t') \rangle = \sigma^2 \delta(t-t')$. Moving to two coupled first-order differential equations, the system is described by
\begin{equation}
\label{eq:appendix_overdamped_EOM}
    \begin{pmatrix}\dot{x}\\\dot{p}\end{pmatrix} = \begin{pmatrix}0 & 1\\
    -\omega_0^2 & -\kappa \end{pmatrix} \begin{pmatrix} x \\ p \end{pmatrix} + \begin{pmatrix} 0 \\ \xi \end{pmatrix}\,.
\end{equation}
Hence, the fluctuation eigenvalues are obtained by diagonalizing the above matrix,
\begin{equation}
    \epsilon^\pm = \frac{1}{2} \left(-\kappa \pm \sqrt{\kappa ^2-4 \omega_0^2} \right)\,,
\end{equation}
where the oscillator is overdamped for $\kappa^2 > 4 \omega_0^2$.
Solving the Fokker-Planck equation corresponding to Eq.~\eqref{eq:appendix_overdamped_EOM} yields the steady state probability distribution $\rho(x,p) = \frac{\kappa  \omega_0}{\pi  \sigma ^2} e^{-\frac{\kappa  \left(p^2+\omega_0^2 x^2\right)}{\sigma ^2}}$. From this, we derive the variance of $x$ and $p$
\begin{align}
    \text{Var}(x) &=\int dx dp \,x^2 \rho(x,p)= \frac{\sigma ^2}{2 \kappa  \omega_0^2}\,,\\
    \text{Var}(p) &=\int dx dp \,p^2 \rho(x,p)= \frac{\sigma ^2}{2 \kappa}\,.
\end{align}

\section{Closed KPO, ground state and diagonalization}
\label{sec:KPO_closed_appendix}

In this Appendix, we provide additional details on the the analysis of the closed KPO [cf.~Eq.~\eqref{eq:KPO_Hamiltonian}], including the mean-field analysis for obtaining the ground state, and the diagonalization of the excitations Hamiltonian on top of the ground state.

We find the ground state of the closed KPO by minimizing the mean-field energy [cf.~Eq.~\eqref{eq:KPO_MF_landscape}] with respect to $\alpha^*$. We obtain region I with the NP (with $\alpha = 0$) as the ground state, if $\text{sign}(U) \Delta < -|G|$, and otherwise the PPS exhibiting a coherent state with $\alpha = \pm |\alpha| e^{i\theta}$ where
\begin{align}
    |\alpha|^2 &= \frac{\Delta + \text{sign}(U) \sqrt{|G|^2}}{U}\,,\\
    \tan(\theta) &= \frac{-G_{\rm Re} - \text{sign}(U) \sqrt{|G|^2}}{G_{\rm Im}}\,.
\end{align}
We now provide the details on the diagonalization procedure applied to the excitations' Hamiltonian Eq.~\eqref{eq:KPO_Hamiltonian}. We start by expanding around the mean-field ground state of interest, i.e., $a = \alpha + c$, where $c$ is now a bosonic operator describing excitations. We can therefore write the KPO Hamiltonian up to second order in the excitations as
\begin{align}
	H_{\mathcal{K}} & = \frac{1}{2}\begin{pmatrix}
					c^\dagger   &   c
				\end{pmatrix}\begin{pmatrix}
					-\Delta + 2U|\alpha|^2& G + U\alpha^2 \\ G^* + U{\alpha^*}^2 & -\Delta + 2U|\alpha|^2
				\end{pmatrix}\begin{pmatrix}
					c   \\  c^\dagger
				\end{pmatrix} \notag \\
				& \qquad + \frac{\Delta - 2U|\alpha|^2}{2}\,.
\end{align}
The corresponding dynamical matrix is
\begin{equation}
    D_{\mathcal{K}} = \frac{1}{2} \begin{pmatrix}
					-\Delta + 2U|\alpha|^2& G + U\alpha^2 \\ -G^* - U{\alpha^*}^2 & \Delta - 2U|\alpha|^2
				\end{pmatrix}\,.
\end{equation}
We solve the eigenvalue problem $D_{\mathcal{K}} \mathbf{v} = \omega \mathbf{v}$ and obtain the eigenfrequencies
\begin{align}
    \omega_{\mathcal{K}}^\pm = & \pm\frac{1}{2} [\Delta^2 -|G|^2 -4U\Delta|\alpha|^2 - 4\alphare\alphaim G_\im U \notag \\
    & - 2 (\alphare^2-\alphaim^2)G_\re U + 3|\alpha|^4 U^2]^{1/2}\,,
\end{align}
the corresponding eigenvectors
\begin{equation}
    \mathbf{v}_\pm = \begin{pmatrix} \frac{G + U\alpha^2}{\omega_{\mathcal{K}}^\pm +\Delta - 2U|\alpha|^2} \\ 1 \end{pmatrix}\,,
\end{equation}
and their associated symplectic norms [cf.~Eq.~\eqref{eq:symplectic_norm}]
\begin{equation}
    ds_\pm^2 = |\frac{G + U\alpha^2}{\omega_{\mathcal{K}}^\pm +\Delta - 2U|\alpha|^2}|^2 - 1\,.
\end{equation}
The Bogoliubov transformation matrix, $V$, has the eigenvectors as columns, ordered placing first all the positive symplectic norm eigenvectors followed by the respective negative symplectic norm ones~\cite{Soriente_2020,Xiao_2009}. In the KPO case we have $V = [\mathbf{v}_+\,\mathbf{v}_-]$ if $ds_+^2>0$ and $ds_-^2<0$. If instead $ds_+^2<0$ and $ds_-^2>0$, then $V = [\mathbf{v}_-\,\mathbf{v}_+]$ . Applying the analysis presented here for the NP, with $\alpha = 0$, we obtain Eqs.~\eqref{eq:KPO_NP_Hamiltonian},\eqref{eq:KPO_NP_eigenfrequencies}, and~\eqref{eq:KPO_symplectic_norm} in the main text.

\section{Open KPO, steady-state and stability}
\label{sec:KPO_open_appendix}

In this Appendix, we present the analytic mean-field steady-state solution for the open KPO [cf.~Eq.~\eqref{eq:KPO_master_eq}], and we study the stability of the solutions to fluctuations. We first solve Eq.~\eqref{eq:KPO_EOM_a} obtaining up to five solutions, including the NP with $\alpha_0=0$, and the PPSs with $\alpha_i  = |\alpha_i| e^{i\theta_i}$, where 
\begin{align}
    |\alpha_1|^2 &= \frac{\Delta - \sqrt{|G|^2 - \kappa^2}}{U}\,,\\
    |\alpha_3|^2 &= \frac{\Delta + \sqrt{|G|^2 - \kappa^2}}{U}\,,\\
    \tan(\theta_1) &= \frac{-G_{\rm Re} + \sqrt{|G|^2 - \kappa^2}}{G_{\rm Im} + \kappa}\, ,\\
    \tan(\theta_3) &= \frac{-G_{\rm Re} - \sqrt{|G|^2 - \kappa^2}}{G_{\rm Im} + \kappa}\, ,
\end{align}
as well as $\alpha_2 = -\alpha_1$ and $\alpha_4 = -\alpha_3$. Requiring that $|\alpha_i|^2\ge 0$ defines the phase boundaries for the existence of the PPS, see Fig.~\ref{fig:KPO_open}(c).

To study the stability of the various solutions, we split Eq.~\eqref{eq:KPO_master_eq} into its real and imaginary parts and expand around the steady-state solutions as outlined in Sec.~\ref{subsec:open_and_steady_states}. Here we generally write $\alpha = \alpha_i$ for the solutions with $i=0,1,2,3,4$. We obtain
\begin{widetext}
\begin{equation}
\label{eq_sm:Jacobian}
    \frac{d}{dt}\begin{pmatrix} \alphare \\ \alphaim \end{pmatrix} = \begin{pmatrix} 2U\alphare\alphaim + G_\im - \kappa & - \Delta + U( - \alphare^2 +\alphaim^2) - G_\re \\
    \Delta - U( \alphare^2 -\alphaim^2) - G_\re & - 2U\alphare\alphaim - G_\im - \kappa \end{pmatrix} \begin{pmatrix} \alphare \\ \alphaim \end{pmatrix}\,,
\end{equation}
to then find the steady-state fluctuation eigenvalues
\begin{equation}
\label{eq:EVKPO}
    \epsilon_{\mathcal{K}}^\pm = -\kappa \pm\sqrt{-[\Delta^2 -|G|^2 -4U\Delta|\alpha|^2 - 4\alphare\alphaim G_\im U - 2 (\alphare^2-\alphaim^2)G_\re U + 3|\alpha|^4 U^2]}\,.
\end{equation}
\end{widetext}
Studying the real parts of $\epsilon_{\mathcal{K}}^\pm$ for the different $\alpha_i$ and assuming $U>0$, we identify the $\alpha_{3,4}$ pair as the stable PPSs, and the $\alpha_{1,2}$ one as the unstable PPSs, dubbed PPS\textsubscript{st} and PPS\textsubscript{un} in the main text, respectively. Note that for $U<0$, the $\alpha_{3,4}$ pair is unstable and the $\alpha_{1,2}$ one is stable.

The variance of fluctuations is calculated from the correlation functions [cf.~Eq.~\eqref{eq:variance_gen}] as outlined in Sec.~\ref{subsec:open_and_steady_states}. We solve the resulting coupled equations for the NP solution in the linear regime ($U\approx0$) and obtain
\begin{align}
    \text{Var}(\alphare) &= \frac{ \left(\Delta ^2+\kappa  (G_{\rm Im}+\kappa )+\Delta  G_{\rm Re}\right)}{ \left(\Delta ^2-G_{\rm Im}^2-G_{\rm Re}^2+\kappa ^2\right)}\,,\label{eq:varKPO1}\\
    \text{Var}(\alphaim) &= \frac{ \left(\Delta ^2+\kappa  (\kappa -G_{\rm Im})-\Delta  G_{\rm Re}\right)}{  \left(\Delta ^2-G_{\rm Im}^2-G_{\rm Re}^2+\kappa ^2\right)}\,.\label{eq:varKPO2}
\end{align}
Crucially, the result discontinuous only at the phase boundary $\Delta = \pm \sqrt{G_{\rm Re}^2 + G_{\rm Im}^2 - \kappa^2}$. Thus, the abrupt splitting in the fluctuation eigenvalues (exceptional point-like scenario) does not lead to any abrupt changes in the variance.

\end{document}